\documentclass[english,aps,prc,twocolumn,floatfix]{revtex4-1}
\usepackage[T1]{fontenc}
\usepackage[latin9]{inputenc}
\usepackage{textcomp}
\usepackage{amstext}
\usepackage{amssymb}
\usepackage{graphicx}
\usepackage{esint}
\def \beq {\begin{equation}}
\def \eeq {\end{equation}}
\def \beqa {\begin{eqnarray}}
\def \eeqa {\end{eqnarray}}

\newcommand{\vect}[1]              
           {\mbox{\boldmath$#1$}}  
\usepackage{babel}
\begin{document}

\preprint{This line only printed with preprint option}

\title{A new Kohn-Sham density functional
based on microscopic nuclear and neutron matter equations of state.}

\author{M. Baldo}

\email{baldo@ct.infn.it}

\affiliation{Instituto Nazionale di Fisica Nucleare, Sezione di Catania, 
Via Santa Sofia 64, I-95123 Catania, Italy}

\author{L.M.Robledo}

\email{luis.robledo@uam.es}

\affiliation{Dep. F\'\i sica Te\'orica (M\'odulo 15), 
Universidad Aut\'onoma de Madrid,
E-28049 Madrid, Spain}

\author{P. Schuck}

\email{schuck@ipno.in2p3.fr}

\affiliation{Institut de Physique Nucl\'eaire, CNRS, UMR8608, F-91406 Orsay, France}
\affiliation{Universit\'e Paris-Sud, Orsay F-91505, France}
\affiliation{ Laboratoire de Physique et Mod\'elisation des Milieux Condens\'es, 
CNRS et Universit\'e Joseph Fourier, 25 Av. des Martyrs, BP 166, 
F-38042 Grenoble Cedex 9, France}

\author{X. Vi\~nas}

\email{xavier@ecm.ub.es}

\affiliation{Departament d'Estructura i Constituents de la Mat\`eria and Institut
de Ci\`encies del Cosmos, Facultat de F\'\i sica, Universitat de Barcelona,
Diagonal \emph{647}, 08028 Barcelona, Spain}


\begin{abstract}

A new version of the Barcelona-Catania-Paris (BCP) energy 
functional is applied to a study of
nuclear masses and other properties. The functional is largely based
on calculated ab initio nuclear and neutron matter equations of state.
Compared to typical Skyrme functionals having 10-12 parameters apart from
spin-orbit and pairing terms, the new functional has only 2 or 3
adjusted parameters, fine tuning the nuclear matter binding energy and
fixing the surface energy of finite nuclei. An energy rms value of 1.58 MeV
is obtained from a fit of these three parameters to the 579 measured
masses reported in the Audi and Wapstra 2003 compilation. This rms
value compares favorably with the one obtained using other successful
mean field theories, which range from 1.5 - 3.0 MeV for optimized Skyrme
functionals and 0.7 - 3.0 for the Gogny functionals. The other properties that
have been calculated and compared to experiment are nuclear radii, the
giant monopole resonance, and spontaneous fission lifetimes.

\end{abstract}
\maketitle

\section{Introduction}

The physics of atomic nuclei and nuclear systems is very rich but 
extremely complicated to describe. It unites many of the toughest 
aspects of the quantum many body problem in a single spot. To name 
only a few, we can mention the existence of four different types of 
fermions (proton/neutron-spin up/down), the relevance of pairing and 
quartet correlations in the dynamics, or the role of  the mean field 
approach as a first order approximation while quantum fluctuations 
are very important as a consequence of the smallness of nuclei. 
Moreover, the  nucleon-nucleon (NN) force is very complicated with 
its hard core, tensor (dipolar) components and strong spin orbit 
interactions. It is remarkable that nuclear theory has nowadays 
mastered a great deal of those challenges and many of the 
experimental data are very well described either on a 
phenomenological basis, or, more rarely, with microscopic 
approaches. In this work we try to progress with the latter strategy. 

Theories aiming towards a global description of low energy nuclear 
data always start with a mean field description and a 
semi-phenomenological nuclear energy density functional. In a recent 
paper \cite{Baldo.08}, we proposed an approach  slightly different 
from the usual for the establishment of a nuclear energy density 
functional. We tried  to follow as much as possible the Kohn-Sham (KS) \cite{KS}
Density Functional Theory (DFT), which is based on the Hohenberg-Kohn 
theorem \cite{HK}, and  is of common use, e.g., in 
atomic and molecular physic \cite {Jon89,Esc96,Per01,Mat02,Tao03,Tao05}. 
The KS method  introduces a set of 
auxiliary single-particle states and takes for the kinetic energy 
the Slater form in this basis. In the KS method the energy density 
functional is written as a sum of two terms. One of them corresponds 
to the uncorrelated kinetic energy density plus the direct 
contribution derived from the underlying two body (Coulomb) 
interaction. The other piece contains an unknown term for the 
exchange and correlation energies. For a practical implementation of 
the method, the latter term has to be guessed and therefore many 
strategies have been developed along the years to describe different 
physical phenomena. A popular strategy in atomic and molecular 
physics is to use accurate theoretical calculations (mainly by 
Monte-Carlo techniques) as a function of the density in simplified 
models where finite size effects are neglected. The results of those 
calculations are used to make an educated guess of the unknown 
exchange and correlation part of the functional in the real system. 
In this way the minimization procedure gives Hartree-like equations 
for the single-particle states, where the potential part includes 
in an effective way the overall exchange and correlation contribution.

The application of the KS-DFT scheme to self-bound systems
like atomic nuclei is not straightforward, because of the absence of
a external confining potential. The usual procedure used with Skyrme functionals
has been justified recently 
\cite{Messud.09,Engel.07,Barnea.07} by reformulating the problem in 
the ``intrinsic" system.

In nuclear physics to build up the exchange and correlation 
contribution amounts to a detailed determination of the ground state 
of symmetric nuclear matter and of neutron matter to be used as a 
guide for finite nuclei energy density functionals. Accurate nuclear 
matter calculations are very complicated and not as much advanced as 
for electron systems of condensed matter. Nonetheless, in recent 
years, quite some progress has been achieved and we will build our 
KS-DFT approach on the equation of state (EOS) of Baldo et al., 
developed in Refs.  \cite{AA,hd1,hd2,tbf1,tbf2,Baldo.04,Baldo.07}, 
which is based on the well known hole line expansion. This EOS is 
calculated from realistic two body and three body interactions 
treated at the Brueckner Hartree-Fock level. It is compatible with 
phenomenological constraints coming from data of heavy ion reactions 
\cite{Baldo.07} as well as from the analysis of astrophysical 
observations \cite {Baldo.12,tara.13}. It is worth mentioning that only two
works exist which tried to follow the same strategy as here, one by 
Fayans \cite{Fayans.98} and the other by Steiner et al \cite{Steiner.04}. 
However, they are based on a somewhat older EOS \cite{FPWFF}.

The functional obtained will be used to describe finite nuclei at 
the mean field level concentrating on the binding energy systematic. 
Therefore, it is important to include beyond mean field effects having a strong 
impact on binding energies and not present in nuclear matter.
The most obvious one is the correlation energy associated to 
symmetry restoration that will be accounted for in the case of 
translational and rotational invariance.

A different approach to nuclear energy density functional (EDF) 
was developed on the basis of the chiral two- and three body 
forces \cite{Hol10,Hol11,Kai12,Geb10,Sto10}. 
The final form of the EDF contains 
a set of contact interactions, similar to the Skyrme functionals, 
with a corresponding set of about a dozen of parameters. This part 
is supplemented by a contribution coming from pion exchanges, which 
is non-local and treated through the density matrix expansion 
procedure. Preliminary calculations \cite{Sto10} on a selected set 
of nuclei show a slight improvement with respect to the standard 
Skyrme functionals and open a promising prospect for large-scale 
fitting procedures.

The reason for such an intense focus on an accurate theoretical 
description of bulk properties of nuclei all over the Nuclide Chart 
is discussed for instance in \cite{Erl12}: many physical scenarios 
involve neutron rich nuclei which are far from reach from an 
experimental point of view and therefore reliable theoretical 
predictions are the only possible option at present. The present 
status is that good agreement between different theoretical models 
is to be expected as long as we do not move towards the neutron drip 
line, a region of relevance in stellar nucleo-synthesis processes. 
Therefore, a new model based on somewhat different ideas closer to 
the KS-DFT approach than the more traditional Skyrme-like EDF 
(inspired by contact central potentials), can help to clarify 
the uncertainties associated to present-day models.

In the present work, a KS approach based on the same microscopic 
bulk input as in \cite {Baldo.08} will be used for the particle- 
hole channel. However, we will be able to reduce the number of 
parameters by two without losing accuracy. As before, we will 
consider an additional bulk parameter for a precise adjustment of 
the numerically obtained E/A value. However, for the surface, we 
will reduce the number of parameters from three to one by the 
condition that, in infinite symmetric and infinite pure neutron 
matter, the strength parameters of the finite range term reproduce 
the coefficients of the quadratic terms of the bulk energy density 
obtained from the microscopic calculation. Therefore, no subtraction 
term in the finite range part as in \cite{Baldo.08} is necessary. 
The range $r_0$ of the surface term remains the only adjustable 
parameter, see the more detailed discussion below. Although there 
are, together with the  strength of the spin-orbit term, three 
parameters in the particle-hole channel, we want to emphasize that 
the most relevant are the ones from bulk and surface. We stress this 
point, because it seems to us a reduction to two basic physical 
inputs to the binding energy of nuclei: energy per particle of 
infinite matter and surface energy! That this is possible is as much 
surprising as it is gratifying! It also should be mentioned that the 
adjustment of those two parameters is extremely sensitive: both have 
to be fine tuned to the order of $10^{-3}$ ! This sensitivity points 
to a well defined physical content of the parameters.

It should be pointed out, however, that KS-DFT addresses only the 
ground state and is, in principle, not tailored to describe 
excited states of the system. Nevertheless, we will also 
use it for the description of the Giant Monopole 
Resonances (GMR's). 

The paper is organized as follows: The following section is devoted 
to briefly recall our previous BCP functional. In the third section, 
the new energy density functional built up in this paper is 
discussed. The results obtained with this improved functional, that 
we will call BCPM (Barcelona-Catania-Paris-Madrid), are also 
presented in the same section. The predictive power of the BCPM 
functional regarding other observable such as charge radii, 
quadrupole and octupole deformations and fission barriers are 
discussed in the fourth section. The ability of the BCPM functional 
for describing the isoscalar giant monopole resonance 
is the subject of the fifth section. Finally, the summary and 
conclusions are given in the last section.

\section{Former BCP functional}

The former BCP functional was proposed in Ref.  \cite{Baldo.08}. This and
subsequent refinements \cite{JPG} are based on the Kohn-Sham Density 
Functional Theory (KS-DFT) where the one body density $\rho ({\bf r})$ 
plays a central role.
In the KS-DFT theory an auxiliary set of $A$ orthonormal wave functions 
$\varphi_i({\bf r})$, where $A$ is the mass number, is introduced to 
express formally the density as if it were obtained from a Slater 
determinant as a sum of the product of single particle wave 
functions 
$$\rho({\bf r}) = \sum_i |\varphi_i({\bf r})|^2,
$$
with the $\varphi$'s determined from the minimization of the ground 
state energy. In condensed matter and atomic physics the energy 
density functional $E[\rho({\bf r})]$ is usually split into two 
parts \cite{KS}:
$$E = T_0[\rho] + W[\rho].$$ 
The first piece  $T_0[\rho]$ corresponds to the uncorrelated kinetic energy written 
in the usual manner as 
$$T_0= \frac{\hbar^2}{2m} \sum_i \int d{\bf r} 
\vert \nabla \varphi_i({\bf r})\vert^2.$$
The other piece, $W[\rho]$, contains the potential energy and the
correlated part of the kinetic  energy. In the BCP family of functionals
this part is given as  the sum of  spin-orbit energy, the 
Coulomb part and the nuclear energy term which depends on the 
neutron and proton densities $\rho_n$ and $\rho_p$, respectively. 
In nuclear physics, contrary to the situation in condensed matter 
and atomic physics, the contribution of the spin-orbit interaction 
to the energy functional is very important. Non-local contributions 
have been included in DFT in several ways long ago (see Ref. 
\cite{Eng03} for a review of this topic). In line with this, we split the 
spin-orbit part into an uncorrelated part $E^{s.o.}$ plus a 
remainder not treated explicitly here. The form of the uncorrelated 
spin-orbit part is taken 
exactly as in typical Skyrme \cite{SLy4,Gor.09} or Gogny 
forces \cite{gog,D1S}. 
Another piece that we explicitly split off is the Coulomb part $E_C$. 
We treat this contribution at lowest order, i.e. the direct term 
plus the exchange contribution in the Slater approximation, that is 
$$
E_C^H= \frac{1}{2} \iint \! d{\bf r}d{\bf r}' \rho_p({\bf r})|{\bf r}-{\bf r'}|^{-1}
\rho_p({\bf r'}),
$$
and
$$
E_C^{ex} = -\frac{3}{4}\left(\frac{3}{\pi}\right)^{1/3} \int \! d{\bf r} {\rho_p({\bf r})}^{4/3}
$$
with $E_C=E_C^{H} + E_C^{ex}$. 

The nuclear energy functional contribution $E_ {int}[\rho_n,\rho_p]$ 
contains the nuclear potential energy as well as additional 
correlations. This contribution is divided into a finite range term 
$E_ {int}^{FR}[\rho_n,\rho_p]$ to account for correct surface 
properties and a bulk correlation part 
$E_{int}^{\infty}[\rho_n,\rho_p]$ that we take from one of the most 
advanced microscopically determined EOS existing so far in the 
literature \cite{AA,hd1,hd2,tbf1,tbf2,Baldo.04,Baldo.07} as 
mentioned before. Collecting all these contributions our final 
KS-DFT  functional reads 
\begin{equation} 
E = T_0 + E^{s.o.} + E_{int}^{\infty} + E_{int}^{FR} + E_C. 
\label{eq6} 
\end{equation}
The functional is supplemented by a density dependent pairing interaction 
and some beyond mean field corrections (see below).

One of the prominent results of our recent work within the KS-DFT 
BCP density functional scheme \cite{Baldo.08} was that, with a very 
reduced number of adjustable parameters to finite nuclei, rms values for binding 
energies and radii came out to be of the same quality as the ones of 
some Skyrme \cite{SLy4} or Gogny \cite{D1S} functionals. The reduction 
of the number of parameters 
stemmed from the fact that, prior to a fit of the functional to 
nuclear masses, the bulk part of the functional $E_{int}^{\infty}$ 
was adjusted to the  microscopic EOS obtained in Refs. \cite
{AA,hd1,hd2,tbf1,tbf2,Baldo.04,Baldo.07} by fitting polynomials in 
the total density $\rho= \rho_n + \rho_p$ to microscopic results in 
symmetric and pure neutron matter, followed by a quadratic 
interpolation in the asymmetry parameter $\beta=(\rho_n - 
\rho_p)/\rho$ between these two limits. 

The curve obtained in the polynomial fit can be slightly changed 
without changing significantly the mean square deviation to the 
microscopic EOS. On the other hand, although the microscopic EOS is 
state-of-the-art, the uncertainties in the underlying interaction and 
the ones attributed to the approximations involved translate to 
minor uncertainties in the computed points of the microscopic EOS 
that will eventually lead to minor changes in the polynomial curve.  
We will exploit this limited freedom to produce polynomial fits with 
slightly different values of E/A at saturation (around E/A=16 MeV). 
The resulting EDFs are then used in the finite nuclei fit (see 
below) and the one leading to the smallest binding energy rms is 
chosen. Only changes of the order of a few hundred of KeV in E/A are 
considered. 

The additional surface term in \cite{Baldo.08} was of the form 
\begin{eqnarray} \label{eq:surfE}
E_{int}^{FR}[\rho_{n},\rho_{p}]  = & \frac{1}{2} & \sum_{t,t'}\!\!\iint \!\! d {\bf r} d {\bf r}'\rho_{t}({\bf 
r})v_{t,t'}({\bf r}-{\bf r}')\rho_{t'}({\bf r}') \\ \nonumber
& - &\frac{1}{2}\sum_{t,t'}\gamma_{t,t'}\int d {\bf r}\rho_{t}({\bf r})\rho_{t'}({\bf r})
\end{eqnarray}
where the index $t$ is the label for neutron and proton, i.e. $t = n,p$, and $\gamma_{t,t'}$ the volume integral
of $v_{t,t'}(r)$. The subtraction term was introduced not to contaminate
the bulk part, determined from the microscopic infinite matter calculation.
For the finite range form factor $v_{t,t'}(r)$ a simple Gaussian
ansatz 
$$
v_{t,t'}(r)=V_{t,t'}e^{-r^{2}/r_{0}^{2}}
$$ 
was made. The strength parameters were chosen to distinguish only 
between like and unlike particles, i.e. 
$$
V_{p,p}=V_{n,n}=V_{L};V_{n,p}=V_{p,n}=V_{U},
$$
so that the finite range term contained three adjustable parameters: 
$V_{L},V_{U}$, and $r_{0}$. Together with the adjustment of the bulk 
and the spin-orbit strength, the functional in \cite{Baldo.08} has 
all in all five adjustable parameters, even though they are not all 
to be considered on the same level of significance. We will come 
back to these considerations later in the text where we 
also will show how to avoid the subtraction term in the finite range 
part of the energy.

In order to describe properly open shell nuclei one has to introduce pairing 
correlations. The formal way of including  pairing within the DFT is 
also known since long from the generalization of the 
Hohenberg-Kohn theorem to paired systems \cite{oli88}. In \cite
{Baldo.08} we  have  introduced pairing in our functional through 
the BCS approach using a zero-range density dependent interaction 
adjusted to reproduce the neutron gaps of the Gogny force in symmetric 
nuclear matter with $m^*=m$ \cite{gar99}.

The free parameters of the energy density functional reported in 
\cite{Baldo.08} were fitted either to reproduce the binding energies 
of a few spherical nuclei or  binding energies and charge radii of 
the same set of spherical nuclei. We called these functionals BCP1 
or BCP2 (Barcelona-Catania-Paris), respectively. The predictive 
power of these  functionals was illustrated by computing the binding 
energies and charge radii of 161 spherical nuclei between $^{16}$Ne 
and $^{224}$U. We found that the rms in energy and charge radii are 
similar to the ones obtained with successful mean-field models such 
us the Gogny D1S force \cite{D1S}, the Skyrme SLy4 interaction \cite
{SLy4} and the relativistic mean field parametrization NL3 \cite{NL3}.

In subsequent publications we explored the properties of BCP1 and
BCP2 in describing quadrupole and octupole deformation ground state
properties, fission, excited octupole states, etc 
\cite{Robledo.08,Vinas.09,Robledo.10}. In dealing with deformed nuclei
the most relevant beyond mean field effect, namely the rotational energy
correction has been considered in an approximate way \cite{egi04}. Also some other
parametrizations of the functional based on fitting to the binding energy
of deformed nuclei  instead of the fitting to spherical
nuclei, have been considered \cite{JPG}. 

We would like to point out that for other energy density 
functionals, like the ones of the Skyrme, Gogny, or relativistic 
mean field type, the number of parameters seems much higher, 
typically more than ten. However, many of those parameters are 
implicitly used to get a reasonable nuclear matter and neutron 
matter equation of state. The advantage of our KS-DFT procedure is 
that one clearly separates the tasks of reproducing the  nuclear 
matter EOS from the most prominent finite size effects, namely the 
surface and Coulomb energies and the spin-orbit potential.

\section{Improvements over BCP and results}

\subsection{New polynomial fitting}

Following \cite{Baldo.08,JPG} we write the bulk part of the energy 
density functional as
\begin{equation}
E_{int}^{\infty}[\rho_{p},\rho_{n}]=\int\! d{\bf r} \big[P_{s}(\rho)(1-\beta^{2})
+P_{n}(\rho)\beta^{2}\big]\rho\label{eq:eq4a},
\end{equation}
where the interpolating polynomials for symmetric and pure neutron matter, 
$P_{s}(\rho)$ and $P_{n}(\rho)$ respectively, read
\begin{equation}
P_s(\rho) = \sum_{n=1}^5 a_n \bigg(\frac{\rho}{\rho_0}\bigg)^n
\qquad
P_n(\rho) = \sum_{n=1}^5 b_n \bigg(\frac{\rho}{\rho_{0n}}\bigg)^n,
\label{eq:polfit}
\end{equation}
where the reference densities are $\rho_0$=0.16 fm$^{-3}$
and $\rho_{0n}$=0.155 fm$^{-3}$, respectively.

As compared to the former version of the BCP energy density 
functional, the fit of the microscopic EOS has been redone in order 
to avoid some rather strong oscillatory behavior  in the 
density-dependent incompressibility of the nuclear matter (see 
Section IV below) when plotted as a function of density. To cure 
this unwanted effect, the number of theoretical points is increased 
and a fifth  order polynomial in the density is chosen 
for fitting the microscopic EOS in symmetric nuclear matter in a 
wider range of densities, up to 0.625 fm$^{-3}$, instead of the 
prescription given in \cite{Baldo.08}, see also \cite{JPG}. 
Furthermore special care has been paid to the smoothness not only of 
the fitted EOS but also of the density-dependent incompressibility. 
This has been achieved by considering only polynomial fits with
smooth high order derivatives with respect to the density.

\begin{figure}
\includegraphics[width=\columnwidth]{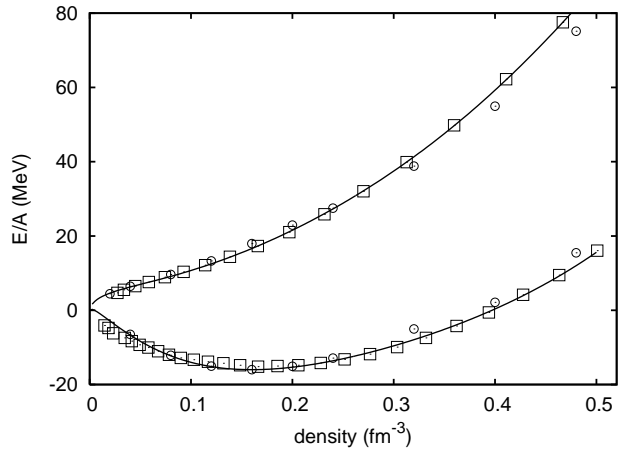}
\includegraphics[width=0.8\columnwidth,angle=-90]{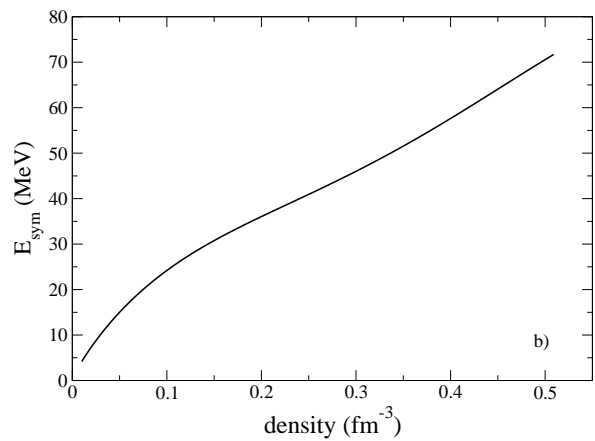}
\caption{Upper panel: EOS of symmetric and neutron matter obtained 
by the microscopic calculation (squares) and the corresponding 
polynomial fits (solid lines). For comparison the microscopic EOS of 
Refs. \cite{AP} are also displayed by open circles. Lower panel: 
Symmetry energy obtained from the polynomial fits}
\label{fig:EOS}
\end{figure}

In Fig. \ref{fig:EOS} the microscopic EOS for nuclear and neutron 
matter as well as the corresponding polynomial fits are shown as 
function of the density. In the lower panel, the symmetry energy 
computed using the polynomial fits as a function of the density is 
also displayed. As discussed in \cite{Mar07}, the low density 
behavior of the microscopic nuclear matter EOS has a characteristic 
trend, usually not reproduced by Skyrme and Gogny functionals ( see 
also ref. \cite{Baldo.04,JPG}), missing there quite a substantial 
part of binding.

Since we want to construct the KS-DFT functional on the basis of the 
microscopic calculations, the bulk part $E_{int}^{\infty}$ of the 
functional, directly related to a realistic EoS, 
is determined once and for all as in Eq. (\ref{eq:eq4a}) together 
with the local density approximation. However, as mentioned before
slightly different polynomial interpolations with different values of
E/A at saturation have been considered to improve the finite nuclei
results for the binding energies. The saturation density has been
kept fixed at the nominal value of 0.16 fm$^{-3}$. For E/A at saturation 
we have explored the interval between 15.97 and 16.03 MeV in steps of 0.01
MeV. For each value of E/A the complete fitting process for finite nuclei has been
performed (see below). The optimal 
value for the saturation energy per particle turns out to be 15.98 MeV. 
The fine tuning of E/A at saturation is to be expected because of the extreme
sensitivity of the mass of heavy nuclei to this parameter. In addition, as the fine
tuning involves finite nuclei calculations where the surface term is
included it can be considered as a way to account for the coupling of volume and surface
in finite nuclei.

\begin{table}
\begin{tabular}{|c|c|c|c|}
\hline 
$n$ & $b_{n}$ & $a_{n}(E/A=16)$ & $a_{n}(E/A=15.98)$\tabularnewline
\hline 
\hline 
1 & -34.972615 & -73.292026 & -73.382673 \tabularnewline
\hline 
2 &  22.182307 &  49.964905 &  50.297798 \tabularnewline
\hline 
3 &  -7.151756 & -18.037601 & -18.366734 \tabularnewline
\hline 
4 &   1.790874 &   3.486176 &   3.608359 \tabularnewline
\hline 
5 &  -0.169591 &  -0.243552 &  -0.258847 \tabularnewline
\hline 
\end{tabular}
\caption{Parameters (in MeV) of the polynomial fits, 
$P_{s}(\rho)=\sum_{n}a_{n}(\rho/\rho_{0})^{n}$
and $P_{n}(\rho)=\sum_{n}b_{n}(\rho/\rho_{0n})^{n}$ for symmetric
and neutron matter, respectively. The reference densities are $\rho_{0}=0.16$
fm$^{-3}$ and $\rho_{0n}=0.155$ fm$^{-3}$. Two sets of parameters
for symmetric matter are given, they correspond to EOS with a minimum
at the given values of $E/A$. The parameters are given with six
digits, enough to obtain binding energies with an accuracy of a couple
of eV. The value $\hbar^{2}c^{2}/2M = 20.75$ MeV has been used.
\label{tab:Polpar}}
\end{table}
 
The values of the coefficients of the fitted polynomials for neutron 
matter and for two choices of symmetric matter leading to slightly 
different values of $E/A$ are reported in Table \ref{tab:Polpar}. A 
nice feature of the coefficients $a_{n}$ is that they follow a 
perfect linear dependence when plotted as a function of $E/A$ and 
therefore the two sets of $a_{n}$ values are enough to reproduce the 
whole range of $E/A$ values considered with only {\it one parameter}.

As an aside, it may be important to mention that the 
density dependent part of the functional depends on integer powers 
of the density only. This is advantageous when trying to overcome 
the self-energy problem that plagues the use of theories beyond the 
mean field based on standard nuclear energy density functionals \cite
{Robledo.07,Robledo.10b}.

\subsection{Infinite matter properties}

The pressure $P$ and density-dependent incompressibility $K$ in 
asymmetric nuclear matter are 
defined in terms of the  energy density in 
asymmetric nuclear matter  ${\cal H}(\rho,\beta)$ as \cite{Khan.10}
\begin{equation}
P = \rho^2 \frac{\partial({\cal H}/\rho)}{\partial \rho} = 
\rho \frac{\partial {\cal H}}{\partial \rho} - {\cal H},
\label{press} 
\end{equation}
and
\begin{equation}
K = 9 \rho \frac{\partial^2 {\cal H}}{\partial \rho^2} =
9 \rho^2 \frac{\partial^2({\cal H}/\rho)}{\partial \rho^2} 
+ \frac{18}{\rho}P
\label{incomp} 
\end{equation}
For symmetric matter ($\beta=0$) at saturation density the
$K(\rho)$ just defined reduces to the well known incompressibility modulus 
$K_0 = 9 \rho^2 \frac{\partial^2({\cal H}/\rho)}{\partial \rho^2}\vert_{\rho=\rho_0}$
where $\rho_0$ is the saturation density.
For later use, it will be convenient to introduce the coefficient $K'$ 
connected to the so-called skewness coefficient \cite{MS} by  
\begin{equation}
K'= -Q = -27 \rho^3 \frac{\partial^3({\cal H}/\rho)}{\partial \rho^3}_{\vert_{\rho=\rho_0}}.
\end{equation}

Other relevant quantities in asymmetric nuclear matter are the 
symmetry energy and its first and second derivatives with respect to 
the density which govern the isovector part of the nuclear 
interaction. The symmetry energy is defined as 

\begin{equation}
E_{sym}(\rho) = \frac{1}{2} \frac{\partial^2}{\partial \beta^2} 
\bigg( \frac{{\cal H}}{\rho}\bigg)_{\vert_{\beta=0}}.
\label{esym} 
\end{equation}
At saturation density one defines the symmetry energy coefficient as 
$J=E_{sym}(\rho_0)$ and two coefficients more, $L$ and $K_{sym}$,
directly related to the first and second derivatives of Eq. (\ref{esym})
with respect to the density at saturation, respectively
\begin{equation}
L = 3\rho \frac{\partial E_{sym}}{\partial \rho}_{\vert_{\rho=\rho_0}}, \qquad
K_{sym} = 9 \rho^2 \frac{\partial^2 E_{sym}}{\partial \rho^2}_{\vert_{\rho=\rho_0}}.
\label{desym} 
\end{equation}
The values of the incompressibility modulus $K_0$ and the 
coefficient $K'$ defined in symmetric nuclear matter as well as the 
coefficients $J$, $L$ and $K_{sym}$ which are related to the 
symmetry energy are displayed in Table \ref{nmatter}. 

The binding energy and the saturation density in nuclear matter are 
constrained by nuclear masses and electron scattering experimental 
data. The range of accepted values of the incompressibility modulus 
$K_0$ is constrained by the experimental excitation energies of the 
isoscalar giant monopole resonance in finite nuclei since the 
pioneering works of Bohigas and collaborators \cite{boh79} and 
Blaizot \cite{Blaizot.80}. However, different estimates of $K_0$ 
using different mean-field models predict slightly different values 
for this coefficient. Recently, a value $K_0 = 230 \pm 30$ MeV has 
been proposed \cite{stone} as a compromise among the different 
available estimates. The value of the symmetry energy coefficient $J$,
that dictates the isospin dependence of the nuclear interaction, is 
constrained by experimental data on heavy-ion collisions, pigmy 
dipole resonances  and analog states. A range $30 \le J \le35$ MeV 
has been recently proposed for this coefficient \cite {Tsang.12}. 
The density content of the symmetry energy is a relevant physical 
quantity related with many phenomena not only in terrestrial nuclei 
but also in neutron stars. Since the celebrated correlation 
established by Brown \cite{Brown.00} between the slope of the 
symmetry energy and the neutron skin in $^{208}$Pb, a considerable 
effort has been devoted to constrain the $L$ parameter from 
available data. Antiprotonic atoms, nuclear masses, heavy-ion 
collisions, giant and pigmy dipole resonances, proton -nucleus 
scattering as well as theoretical calculations using a microscopic 
interactions have been used to extract the $L$ coefficient (see Ref. 
\cite{Warda.09} and references therein). From a compilation of the 
existing data the range of values $L= 55 \pm 25$ MeV has been proposed 
\cite{Warda.09,Centelles.09}. The other two coefficients that 
characterize asymmetric nuclear matter around saturation, namely $K'$
and $K_{sym}$ are more uncertain. An estimate $K' = 700 \pm 500$ 
MeV has been proposed in Ref. \cite{stone}. The curvature of the 
symmetry energy, $K_{sym}$, can be inferred from some 
non-relativistic \cite {Nayak.90} and relativistic \cite{Chossy.97} 
calculations in nuclear matter. From these results a range $-200 \le K_ {sym} \le 150$ 
MeV can be inferred.  As can be seen from a direct inspection of 
Table \ref{nmatter}, the nuclear matter properties of our BCPM 
energy density functional lie within the accepted ranges of values of the 
different nuclear matter quantities.

\begin{table}[ht]
\caption{Infinite nuclear matter
properties of BCPM. All the parameters, except $\rho_0$ (in fm$^{-3}$) and the 
dimensionless effective mass $m/m^*$, are given in MeV}
\begin{center}
\begin{tabular}{ccccccccc}
& $B/A$ & $\rho_0$ & $m/m^*$ & $J$ &
$L$ & $K_{0}$ & $K'$ & $K_{sym}$\\
\hline
& -15.98 & 0.16 & 1.00 & 31.90 & 52.96 & 212.4 & 879.6 & -96.75 \\
\hline
\end{tabular}
\end{center}
\label{nmatter} 
\end{table}

\subsection{Surface term strengths derived from nuclear matter data
and spin-orbit contribution}

In the new version of the functional the Gaussian form factors in the surface
term are free to have different ranges depending on the isospin 
channel $r_{0L}$ and $r_{0U}$ introducing in this way a new parameter as 
compared to the old BCP1. However, the strengths $V_{L}$ and $V_{U}$ are 
now no longer considered as free parameters. They have been determined in such a
way that the bulk limit of the surface term in Eq. (\ref{eq:surfE}), that is
$$
\frac{1}{2}\sum_{t,t'}\gamma_{t,t'}\int d^{3}r\rho_{t}({\bf r})\rho_{t'}(({\bf r}),
$$
reproduces  the $\rho^{2}$ term of the bulk part of the energy density 
in Eq. (\ref{eq:eq4a}). Therefore, no subtraction as in Eq. (\ref{eq:surfE}) 
is required. This procedure imposes the following relationships between 
the surface term strengths and the coefficients $a_{1}$ and $b_{1}$ of 
the polynomial fits of symmetric and neutron matter, respectively
\begin{eqnarray}\label{eq:VU}
V_{L} & = & \frac{2\tilde{b}_{1}}{\pi^{3/2}r_{0L}^{3}\rho_{0}}\label{eq:VL}\\
V_{U} & = & \frac{4a_{1}-2\tilde{b}_{1}}{\pi^{3/2}r_{0U}^{3}\rho_{0}}.
\end{eqnarray}
Here $r_{0L}$ and $r_{0U}$ are the ranges of the Gaussian in the surface term, 
as defined above. The parameter
$\tilde{b}_{1}$, defined as $\tilde{b}_{1}=b_{1} \frac{\rho_{0}}{\rho_{0n}}$,
has also been introduced to take into account the different reference densities
for symmetric and neutron matter.

On the other hand, the spin-orbit strength does not need to be adjusted
because the final result is nearly independent of it. The spin-orbit
strength is fixed by the requirement to reproduce the magic numbers and
therefore depends on the major shells energy separation. The latter is
roughly speaking inversely proportional to the effective mass. The same
dependence with the effective mass should hold for the spin-orbit strength.
Therefore, we will take the value $W_{LS}=90.5$ MeV which is consistent with 
BCPM's effective mass of one and the spin-orbit strength $W_{LS}=130$ MeV 
of the Gogny force ($m^{*}=0.7 m $). 


From the above discussion we conclude that the number of parameters in the 
surface term gets reduced from three to two, namely the $r_{0L}$ and 
$r_{0U}$ ranges. As mentioned, the spin-orbit strength $W_{LS}$ is 
practically fixed from the beginning. On the other hand, as we will see below, 
the actual fit allows for $r_{0L} = r_{0U}=r_0$ and, thus, one is 
left with one  parameter only.

\subsection{Fitting protocol for finite nuclei}

\begin{figure}
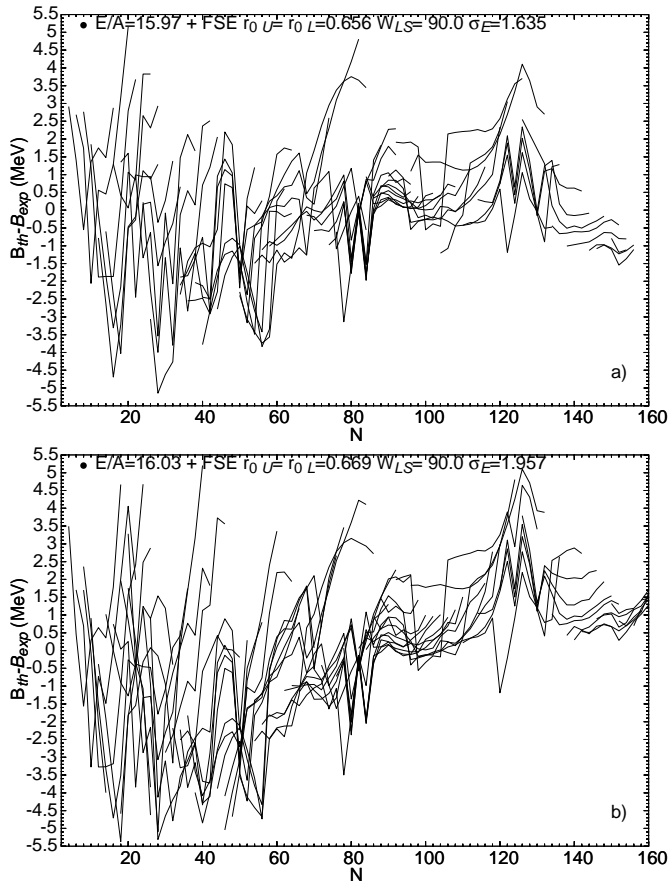

\includegraphics[angle=-90,width=\columnwidth]{DeltaE_15.97_0.656_90.ps}

\includegraphics[angle=-90,width=\columnwidth]{DeltaE_16.03_0.669_90.ps}
\caption{Binding energy differences $\Delta B=B_\textrm{th} -
B_\textrm{Exp}$ (in MeV) as a function of neutron number $N$ for two
choices of $E/A$, namely $E/A=15.97$ in panel a) and $E/A=16.03$ in panel b).}
\label{Fig3} 
\end{figure}

Owing to the simplicity of the BCP
energy density functional we are in the position to perform mass table
evaluations on a personal computer in a reasonable amount of time
( a few hours ). Therefore we have decided to include in the fit not
only the masses of spherical nuclei but also the ones corresponding
to deformed ground states. 
The only limitation has been the restriction
to even-even nuclei. In the fit we have searched for the values
of the parameters that minimize the rms deviation for the binding energies
$$
\sigma^{2} (E) = \frac{1}{N} \sum_{i=1}^{N} (B_{th}(i)-B_{exp}(i))^{2}
$$
where the sum runs over the set of 579 even-even nuclei with known experimental
binding energies, as given in the 2003 evaluation of Audi and Wapstra  \cite{aud03}.
The 193 extrapolated binding 
energies included in Audi and Wapstra's work are excluded from our analysis, although we will use 
them to explore the quality of our fit. 
Finite nuclei properties are computed within the mean field 
framework provided by the well known Hartree-Fock-Bogoliubov (HFB) 
method \cite{RS80}. The pairing interaction is the same as in the 
original formulation of the functional \cite{Baldo.08} but using the 
HFB instead of the BCS method. 

\subsection{Details of the HFB calculations}

The HFB calculations have been restricted to axially symmetric solutions.
The quasiparticle operators of the Bogoliubov canonical transformation 
have been expanded in a harmonic oscillator basis with a size 
(the number of shells) depending on the nuclei considered. 
For nuclei with $Z<50$ we take a basis with eleven harmonic oscillator shells, for 
$50<Z<82$ we consider 13 shells and for $Z>$82 a 15 shell 
basis is used. These bases are used in a set of constrained calculations
providing the potential energy as a function of the quadrupole moment for each
nucleus. The minima of the potential energy are used as starting wave functions 
for an unrestricted minimization. Once the minimum energy solution is found
for each nucleus, an extrapolation to an 
infinite basis is performed in the spirit of Ref. \cite{Hilaire.07}: 
the HFB equation is solved with and enlarged basis containing two extra shells and 
the infinite basis energy is obtained by the extrapolation formula 
$E(\infty)=2E(N+2)-E(N)$ ($N$ is the initial number of shells). 

For the solution of the non-linear HFB 
equation, an approximate second order gradient method is used 
\cite{rob11}. The information about the energy curvature reduces the 
number of iterations substantially as compared to other methods. As 
it is customary in this kind of calculations, the Coulomb exchange 
contribution is computed in the Slater approximation \cite {TS.74} 
and the Coulomb anti-pairing effect is not explicitly considered 
(see \cite{Ang.01} for a discussion of this issue). The two body 
kinetic energy correction, which is typically considered as a way to 
correct for the lack of translational invariance of the whole 
procedure, has been taken into account with the pocket formula of 
Refs. \cite{But.84,Sou.03} as in previous versions of the BCP 
functional. In deformed nuclei the correlation energy stemming from 
symmetry restoration of the rotational invariance (the rotational 
energy correction) plays a relevant role. It rapidly changes from 
magic or semi-magic nuclei, where it is zero or very small, to 
strongly deformed mid-shell nuclei where it reaches values that can 
be as large as 6 or 7 MeV in heavy nuclei. This is a consequence of 
its almost linear dependence with quadrupole deformation and 
therefore its inclusion is relevant to describe masses all over the 
Nuclide Chart. The correct way to compute this quantity is by 
evaluating the angular momentum projected energy associated with 
each intrinsic state. This is a tremendous task that can fortunately 
be alleviated by considering an approximation to the projected 
energy that is obtained in the spirit of the Topological Gaussian 
Overlap Approximation. This fully microscopic formula, which is similar to the 
rotational energy correction, can be easily evaluated at the mean 
field level and does not involve any phenomenological parameter 
(see Ref. \cite{egi04} for details). Finally, as already 
mentioned, for the pairing part we have taken the density dependent  
zero range force of Ref. \cite{gar99} that was devised to mimic the 
behavior of Gogny D1 pairing gap in nuclear matter. We have taken 
care of our effective mass equal to one by renormalizing the pairing 
force strength given in \cite{gar99}.

\subsection{Determination of parameters}

The three initial free parameters (i.e, the two ranges and 
spin-orbit) were at first considered in the fit but soon it became 
clear that the $E/A$ value given by the polynomial fit should also 
be taken into account as a free parameter in the way we discussed 
above. Out of the four, it turns out that $\sigma(E)$ has a very 
smooth dependency on $W_{LS}$ and the minimum value of $\sigma(E)$ 
was always obtained for spin orbit strength values around 90 MeV fm~
$^5$. As explained before, this value is consistent with the value of 130 MeV fm~$^5$ 
used in Gogny D1S (the difference is related to the different 
effective masses which are $m^{*}=m$ for BCP and $m^{*}=0.7 m$ for 
Gogny D1S and therefore $0.7\times 130\approx 90$). Another relevant 
observation is that the binding energy difference $\Delta 
B=B_{\textrm{th}}-B_{\textrm{Exp}}$ shows a linear dependence with 
mass number $A$ (which is sometimes masked by large fluctuations at 
low A) with a slope that is intimately related to the value of $E/A$ 
for the bulk. It is almost zero for $E/A\approx15.97$ and is clearly 
different from zero and positive for $E/A=16.03$ as can be observed 
in Fig. \ref{Fig3}. It turns out that the value $E/A=15.98$ yields 
the lowest $\sigma(E)$ value. The final relevant observation is that 
$\sigma(E)$ depends sensitively on the values of $r_{0L}$ and $r_{0U}$ 
and an accuracy of one part in $5\times10^{3}$ is required to obtain 
reasonable values for that quantity. Systematic explorations with a reduced
set of spherical nuclei shows that there are two 
sets of $r_{0L}$ and $r_{0U}$ values that lead to reasonable values 
of $\sigma(E)$, namely $r_{0L}$ equal $r_{0U}$ with values around  0.660 fm and 
$r_{0L}$ taking values around 0.490 fm and $r_{0U}$ around 1.050 fm. In the latter 
case, the value of $r_{0U}$ is more critical than the $r_{0L}$ value 
and it has to be kept at the value $r_{0U}=1.046$ fm leaving only 
one free parameter to play, namely $r_{0L}$. Although the values of 
$\sigma(E)$ (considering the 579 nuclei) obtained by minimizing with 
respect to $r_{0L}$ and $r_{0U}$ in the neighborhood of the two possibilities 
$r_{0L}=r_{0U}\approx 0.660$ fm and $r_{0L}\approx 0.490$ fm and 
$r_{0U}=1.046$ fm are similar, the first choice produces a  $\Delta B$ 
plot that  looks smoother than the  second and therefore we will from now
on restrict ourselves to the first choice. We have considered $r_{0L} =r_{0U}$
values in the interval between 0.650 fm and 0.670 fm in steps of 0.002 fm 
except for the interval bracketing the minimum where a step of 0.001 fm
has been considered.

After all these considerations, and having carried 
out a large set of mass table calculations to validate our choice, 
we arrive to the conclusion that $E/A=15.98$ MeV, $W_{LS}=90.5$ MeV 
fm~$^5$, and $r_{0U}=r_{0L}=0.659$ fm is the best choice leading to 
a $\sigma(E)$ value of 1.58 MeV for the set of 579 nuclei considered 
in the fit. In the following, as already mentioned,  we will denote 
the new set of parameters as the BCPM functional. 

Regarding the smallness of the range parameter one should 
remember that even for a zero range force the surface energy is not 
zero, due to the surface spread of the quantum single particle wave 
functions. Therefore, it only needs very little extra smearing of 
the density via a finite range force to get to the experimental 
value of the surface energy. To quantify this statement, we have 
calculated the surface energy provided by the BCPM functional using 
the self-consistent Extended Thomas-Fermi approach including 
$\hbar^2$ corrections (ETF-$\hbar^2$) \cite{Centelles.98}. 
According to this reference, the HF surface 
energy can be estimated by renormalizing the Weizs\"acker term in 
the semi-classical kinetic energy density. We have obtained the 
renormalizing factor from a fit to earlier exact HF calculations 
(using BCP1 and BCP2 functionals \cite{Robledo.08}). In this way we 
estimate the HF surface energy associated with the BCPM functional 
to be 17.68 MeV.

\begin{figure}
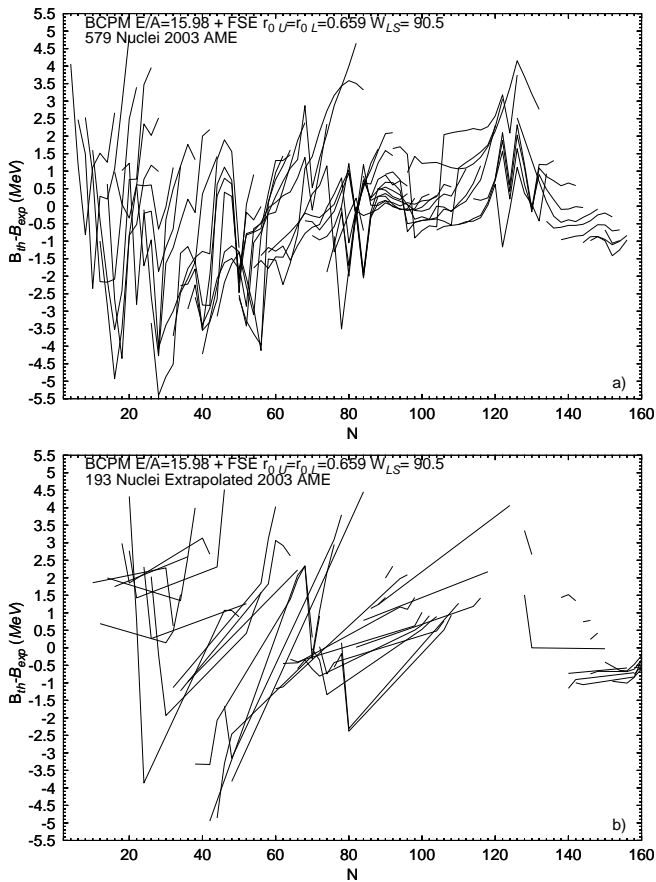

\includegraphics[angle=-90,width=\columnwidth]{DeltaE_0.659.ps}%

\includegraphics[angle=-90,width=\columnwidth]{DeltaE_193Extra}
\caption{In panel a), the binding energy difference $\Delta 
B=B_\textrm{th} - B_\textrm{Exp}$ (in MeV) is plotted with our optimal
set of parameters (BCPM) for the 579 
nuclei of Audi's AME 2003 as a function of neutron number N. Points 
corresponding to the same isotope are connected by straight lines. 
In panel b), the same quantity is plotted but this time for 
the extra set of 193 nuclei in Audi's compilation with 
"experimental" values obtained from extrapolation and/or systematic.
\label{fig:SEFinal}}
\end{figure}

In panel a) of Fig. \ref{fig:SEFinal} we present the $\Delta B$ 
quantities for the $579$ nuclei with measured masses as a function 
of neutron number. The first noticeable conclusion is that, as 
expected, the agreement with experimental data is much better for 
heavy nuclei than for light ones. A more quantitative assessment is 
obtained by looking at $\sigma(E)$ for different sets of nuclei: 
taking into account nuclei with $A>40$ (536 nuclei) the $\sigma(E)$ 
for the energy gets reduced to 1.51 MeV and for the remaining 43 
nuclei with $A<40$ it increases up to 2.31 MeV. When only nuclei 
with $A>80$ (452) are considered a $\sigma(E)$ value of 1.34 MeV is 
obtained. A glance at the plot reveals that those figures are 
consistent: for heavy nuclei the fluctuations around zero are of 
small amplitude whereas for light nuclei the fluctuations lead to 
discrepancies as high as 5.5 MeV in the upper side or -5 MeV in the 
lower one. We also observe strong deviations around magic neutron 
numbers $N=82$ and $N=126$. In the lower panel of Fig. \ref
{fig:SEFinal} we have plotted the $\Delta B$ for those 193 nuclei 
obtained by extrapolation in the  Audi and Wapstra compilation. 
We observe a very good 
reproduction of the binding energies for $N\approx150$ but as N 
decreases strong deviations from the extrapolations are observed. 
For the 193 nuclei we obtain $\sigma(E)=2.34$ MeV which deviates 
quite a lot from the nominal value of 1.58 MeV. However, the value 
for $A>80$ nuclei (146 nuclei) is $1.65$ MeV which is much closer to 
our reference value. For $A<80$ (47 nuclei) we get $\sigma(E)=3.76$ 
MeV. The discrepancies can be partly attributed to deficiencies in the 
extrapolation.

\subsection{Variance of the parameters}

In order to assess in a more quantitative way the variance of the 
parameters we have performed three sets of calculations. In the 
first, the values of $E/A$ and $W_{LS}$ have been kept and different 
choices for $r_{0L}=r_{0L}$ have been made. The results are 
summarized in Table \ref{tab:r0} where we observe how a change of 
$0.003$ fm in the $r_{0U}=r_{0L}$ values (a 0.5 \% change) modifies 
the $\sigma(E)$ values by more than 40 \% from 1.58 MeV to 2.16 MeV 
for $r_{0U}=r_{0L}=0.662$ fm (2.23 MeV for $r_{0U}=r_{0L}=0.656$ 
fm). It is also worth mentioning that the $\sigma(E)$ value for 
$A<40$ remains essentially constant as a function of $r_{0L}=r_{0U}$ 
indicating a high degree of randomness in $\Delta B$ for light 
nuclei that points to a clear deficiency of the mean field theory to 
describe such light systems. On the other hand, for heavier nuclei 
with $A>40$ there is a definite parabolic structure with the minimum 
centered around $r_{0U}=r_{0L}= 0.659$ fm. For $A \geq 80$ the 
$\sigma (E)$ value at the minimum gets reduced to 1.35 MeV in 
agreement with the smaller values of $\Delta B$ observed for heavy 
nuclei. On the other hand, the $A<80$ $\sigma (E)$ values seem to 
favor smaller values of $r_{0U}=r_{0L}$ but the gain does not seem 
to be very significant. From the values in the Table we can conclude 
that the use of different values of $r_{0U}=r_{0L}$ for different 
regions of the mass table is only marginally beneficial.

\begin{figure}
\includegraphics[width=\columnwidth]{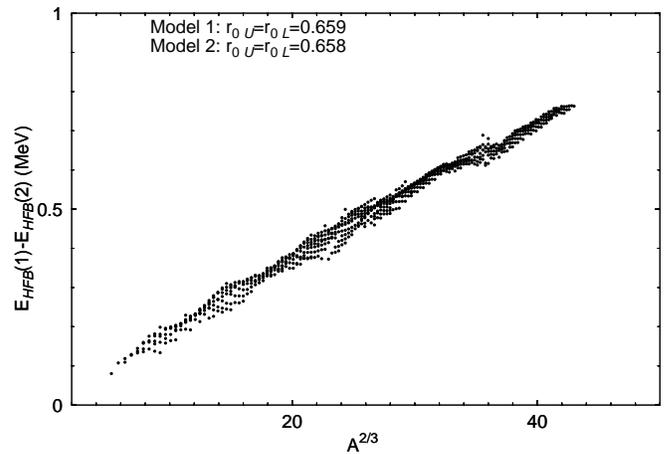}%
\caption{HFB energy difference computed with $r_{0L}=r_{0U}=0.659$ fm and  $r_{0L}=r_{0U}=0.658$ 
as a function of $A^{2/3}$.
\label{fig:DE}}
\end{figure}

The sensitivity of $\sigma (E)$ to the value of $r_{0L}=r_{0U}$ parameter
is a clear indication of the strong connection between this parameter
and the contribution of the surface term to the binding energy. To clearly
establish this connection we have plotted in Fig. \ref{fig:DE} the difference
in the HFB energies computed with $r_{0L}=r_{0U}=0.659$ fm and  $r_{0L}=r_{0U}=0.658$ 
as a function of $A^{2/3}$. The figure shows a clear linear behavior that
shows the connection between $r_{0L}=r_{0U}$ and the surface energy.

The sensitivity of $\sigma (E)$ to $r_{0L}=r_{0U}$ to the level of 0.5\%{}
should not be surprising as it is comparable to the sensitivity of the same quantity to some parameters
of Gogny or Skyrme like functionals. For instance, the strength parameter $t_{3}$
of the density dependent part of the Gogny D1S EDF produces also strong variations
of $\sigma (E)$, a change of 0.1 \% in its value increases the $\sigma (E)$ 
by a factor of 2 (that is a 100 \% change !). This is not a peculiarity of
Gogny D1S as the same sensitivity is observed with Gogny D1M and should be
present in any Skyrme with $t_{3}$ values of the order of $10^{3}$ MeV or
higher.

\begin{table}
\begin{tabular}{|c|c|c|c|c|c|}
\hline 
$r_{0\, L}$ & $\sigma(E)$  & $A\geq40$ & $A<40$ & $A\geq80$ & $A<80$\tabularnewline
\hline 
\hline 
0.656 & 2.23 & 2.22 & 2.35 & 2.27 & 2.11 \tabularnewline
\hline 
0.657 & 1.91 & 1.87 & 2.32 & 1.84 & 2.11 \tabularnewline
\hline 
0.658 & 1.68 & 1.62 & 2.31 & 1.52 & 2.16 \tabularnewline
\hline 
0.659 & 1.58 & 1.51 & 2.31 & 1.35 & 2.23 \tabularnewline
\hline 
0.660 & 1.65 & 1.58 & 2.33 & 1.40 & 2.33 \tabularnewline
\hline 
0.661 & 1.85 & 1.82 & 2.35 & 1.64 & 2.45 \tabularnewline
\hline 
0.662 & 2.16 & 2.14 & 2.40 & 2.02 & 2.61\tabularnewline
\hline 
\end{tabular}\caption{Values of $\sigma(E)$ (MeV) for different values of $r_{0L}$ (fm)
in the $r_{0L}=r_{0L}$ case. The values of $E/A$ and $W_{LS}$ are fixed
at the BCPM values of 15.98 MeV and 90.5 MeV fm~$^5$, respectively. 
Different columns correspond to different choices in the $A$ values included in the
evaluation of  $\sigma(E)$. For $A\geq40$ ($A\geq80$) 536 (452) nuclei are included. \label{tab:r0}}
\end{table}

Keeping $E/A$ constant and equal to its optimal value of 15.98 MeV
and varying $W_{LS}$ with $r_{0L}=r_{0L}$
values determined as to minimize $\sigma(E)$ leads to the results
of Table \ref{tab:SEvsWLS}. There we observe that changes of the
order of 5\% in $W_{LS}$ lead to changes of the order of 6 \% in
$\sigma(E)$. This result clearly indicates that, as expected, the
dependency of the masses on the spin-orbit strength is rather weak.

\begin{table}
\begin{tabular}{|c|c|c|c|c|}
\hline 
$r_{0\, L}$ (fm) & $W_{LS}$  & $\sigma(E)$ & $\sigma(E)$ ($A\geq80$, 452) & $\sigma(E)$ ($A<80$,127) \tabularnewline
\hline 
\hline 
0.653 & 85.5 & 1.62 & 1.38 & 2.27\tabularnewline
\hline 
0.659 & 90.5 & 1.58 & 1.34 & 2.23\tabularnewline
\hline 
0.665 & 95.5 & 1.67 & 1.44 & 2.32\tabularnewline
\hline 
\end{tabular}
\caption{Values of $\sigma(E)$ (in MeV) for different choices of 
$W_{LS}$ (in MeV fm$^{5}$) and at the fixed value $E/A=15.98$ MeV. 
For each $W_{LS}$ the values  $r_{0L}=r_{0L}$ have been fixed to 
minimize $\sigma(E)$. The values in parenthesis following $\sigma(E)$
represent the number of nuclei used in the evaluation of $\sigma(E)$.
\label{tab:SEvsWLS}}
\end{table}

The second derivatives of the $\sigma (E)$ with respect to the
parameters of the fit have been evaluated and their values are given in 
Table \ref{tab:d2se}. By looking at the last row corresponding to the
normalized variables we can read off how a 1\% change in the corresponding
value of the variables is going to modify the $\sigma (E)$ value at the
minimum. For $W_{LS}$ that 1\% change will lead to a change of 0.16 MeV (the smallest),
whereas for $r_{0U}$ it will imply a change of 3.72 MeV (the largest). We
conclude that the quality of our fit is more sensitive to $r_{0U}$ than
to $r_{0L}$. An eigenvalue analysis of the Hessian matrix for $r_{0U}$ and
$r_{0L}$ shows that there is a nearly zero eigenvalue for those combinations
satisfying $r_{0U}=-0.35 r_{0L}$. This implies that $r_{0U}$ and $r_{0L}$
are not independent variables at least around their optimal values 
$r_{0U}=r_{0L}=0.659$. The other eigenvalue is $4.17 \times 10^{4}$ and
corresponds to the combination $r_{0U}= 2.9 r_{0L}$.

Finally, in order to test the sensitivity of our procedure to changes in the 
degree of the polynomial fitting symmetric matter EoS, we have considered two 
alternative fits, one with a fourth order polynomial and another one 
with a sixth order one. The incompressibility coefficient $K_{0}$ in symmetric
nuclear matter takes the value  238.55 
MeV for $n=4$ ($n$ represents here the polynomial order) and 226.50 
MeV in the $n=6$ case. In both cases, the fine tuning of E/A at saturation
gives a  value of 15.98 MeV as in our final choice of n=5 for 
BCPM. The fit to finite nuclei with the corresponding functionals 
leads to slight modifications on the value of $r_0 \equiv r_{0, U} = 
r_{0, L}$ but the $\sigma (E)$ value at the minimum changes little. 
For $n=4$ we get $\sigma (E)=1.69$ MeV for $r_{0} = 0.649 $ fm and for $n=6$ we 
get $\sigma (E)=1.64$ MeV for $r_{0} = 0.654$ fm. Clearly, the performance
of BCPM in reproducing binding energies does not depend in a significant
way on the degree of the polynomial fitting symmetric matter.

\begin{table}
\begin{tabular}{|c|c|c|c|c|c|}
\hline 
 & x=$W_{LS}$  & x=$W_{LS}$ & x=$r_{0U}$ & x=$r_{0L}$ &  x=$r_{0U}$\tabularnewline
 & y=$W_{LS}$ & y=$r_{0L}$ & y=$r_{0U}$ & y=$r_{0L}$ &  y=$r_{0L}$\tabularnewline
\hline 
\hline 
$\frac{\partial^{2} \sigma (E)}{\partial x \partial y}$ & 
0.195 & -166.812 & 8.58$\times 10^{4}$ & 1.11$\times 10^{4}$ & 2.99 $\times 10^{4}$\tabularnewline
\hline 
$\frac{\partial^{2} \sigma (E)}{\partial \bar{x} \partial \bar{y}}$ & 
0.16 $\times 10^{4}$ & -0.99  $\times 10^{4}$ & 3.72$\times 10^{4}$ & 0.48 $\times 10^{4}$ & 1.30$\times 10^{4}$\tabularnewline
\hline 
\end{tabular}
\caption{Values of the second derivative of $\sigma(E)$ (in MeV) with 
respect to the $W_{LS}$ (in MeV fm$^{5}$), $r_{0U}$ (in fm)  and $r_{0L}$
(in fm) parameters. The last row is for the values corresponding
to variables normalized to the value 1 for the parameters that minimize
$\sigma (E)$.\label{tab:d2se}}
\end{table}

\begin{figure*}
\includegraphics[scale=0.85]{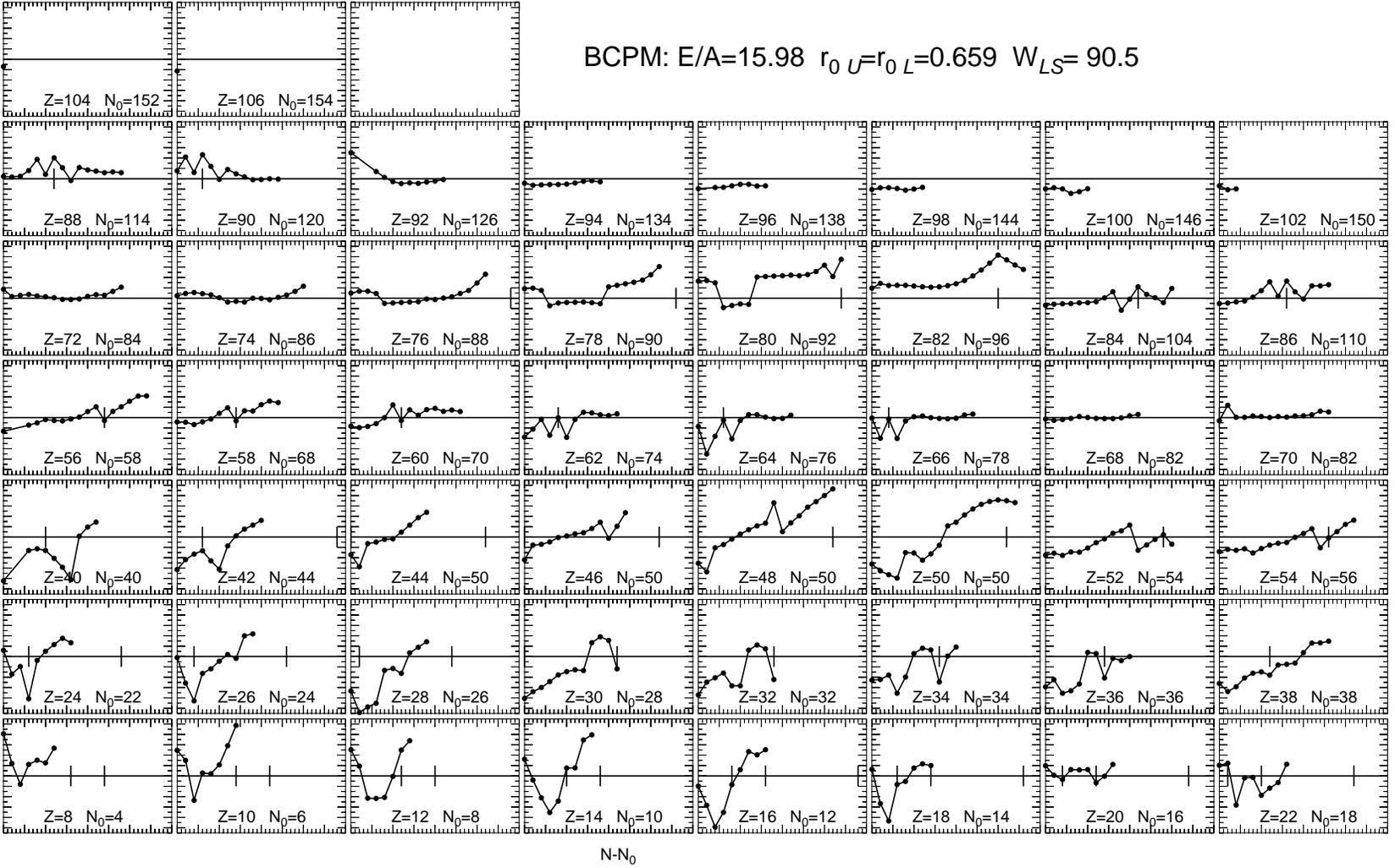}
\caption{The binding energy difference $\Delta B=B_\textrm{th}-B_{\textrm{exp}}$ 
(MeV) is plotted as a function of the shifted (by N$_{0}$) neutron number N-N$_{0}$ 
for all the isotopic chains considered. The values of Z and the neutron number
shift N$_{0}$ for each chain are given in the corresponding panel.
The ordinate $\Delta B$ axis ranges from -5.5 MeV to 5.5 MeV with long ticks every 1 MeV.
The N-N$_{0}$ axis spans a range of 40 units with long ticks every 5 units
and short ones every 1 unit. In every panel, a horizontal 
line corresponding to $\Delta B=0$ has been plotted to 
guide the eye. Additional perpendicular lines signaling the position of 
magic neutron numbers have also been included.}
\label{Fig4} 
\end{figure*}

\subsection{Isotope chain analysis and role of quadrupole deformation}

We would like to discuss the binding energy differences $\Delta B$ 
from a different perspective by looking at its behavior with neutron 
number for each isotopic chain considered. The quantity is plotted 
in Fig. \ref{Fig4}. In each panel, values for the number of 
protons Z and the number of neutrons N$_{0}$ signaling the origin 
of the x axis are given. Perpendicular lines in the $\Delta B=0$ 
horizontal lines indicate the location of magic neutron numbers. 
From this plot, we conclude that for medium mass and heavy nuclei 
the $\Delta B$ curves are rather flat as a function of N  except for 
those Z values corresponding to magic numbers (Z=50 and 82) plus or 
minus two. For lighter nuclei the result is less certain but overall 
we can say that again values of Z of 38, 40 (not magic but a 
sub-shell closure) and Z= 28 and 30 have larger values of $\Delta B$ 
than the other Z values. The impact on the $\sigma(E)$ can be significant. 
For instance, if nuclei with Z=48, 50, 80 and 82 are not 
included in the evaluation of $\sigma(E)$ its value goes down to 
1.46 for 514 nuclei. Partial $\sigma(E)$ values corresponding to 
$A>80$ go from 1.34 (452 nuclei) when all possible Z values are 
considered to 1.09 (387 nuclei) when the Z values indicated above 
are not considered. It is also observed that for neutron numbers 
equal to magic ones plus or minus two the binding energy difference 
$\Delta B$ always show peaks. The effect is reinforced when both 
proton and neutron numbers are magic, in such a way that the two nuclei 
with the largest $\Delta B$ values are $^{208}$Pb (Z=82) and $^{130}$Cd (Z=48). 
Let us finally mention that the peculiarities of this figure are 
also observed for other forces like Gogny D1S \cite
{D1SWEB,Hilaire.07}. The origin of these findings is uncertain but 
probably has to do with dynamical correlations associated to shape 
and pairing degrees of freedom.

Another interesting piece of information is the contribution to 
$\sigma (E)$ according to the deformation of the nuclei. Of the 579 
nuclei considered in the evaluation of $\sigma (E)$ there are 365 
with a ground state deformation parameter $\beta_{2}$ greater, in 
absolute value, than 0.1 (a value that corresponds to a moderately 
deformed system). The $\sigma (E)$ value for those deformed systems 
gets reduced to 1.08 MeV whereas the complementary one corresponding 
to near spherical systems (214 nuclei) grows up to a value of 2.19 
MeV. This result is in the line of recent claims that well deformed 
systems are described better by mean field models than the spherical 
ones \cite{Niksic.08}. The result is also consistent with the 
previous finding concerning the maximum values of $\Delta B$ for 
magic or semi-magic nuclei as those nuclei tend to have a nearly 
spherical ground state shape.

\subsection{Comparison with other approaches}

In Ref \cite{Ber05} the idea of using minimax fitting criteria 
instead of the more traditional rms ones is discussed. The authors considered 
the Chebyshev norm $\epsilon$ which is based on the maximum absolute 
value of the residuals $r_{A}=E_\textrm{data} - E_\textrm{theory}$. 
The object of the fit is to minimize $\epsilon$. The result for SLy4 
yields a value of 4.8 MeV with the six critical nuclei $^{20}$Mg, 
$^{208}$Pb (both over-bound), $^{60}$Zn, $^{64}$Ge and $^{250}$Cm. 
We have not carried out a minimax refit of BCPM but we think the 
values of the largest residuals are worth to be presented. The 
nuclei with the largest negative $r_A$ values are $^{56}$Ni (-5.40 
MeV), $^{32}$S (-4.92 MeV), and some of their isotopes. In the 
positive side, we have $^{30}$Ne (4.87), $^{130}$Cd (4.65) and 
$^{208}$Pb (4.15). In agreement with the results of \cite{Ber05} we 
note that most of the nuclei with large values of $|r_A|$ are magic 
or semi-magic nuclei.

To finish this subsection we would like to compare our results with 
those obtained with the different parametrizations of the Gogny 
force, namely D1S, D1N \cite{D1N} and D1M \cite{D1M}. The results 
for $\sigma (E)$ and the three parametrizations are given in Table 
\ref{tab:Gogny}. For the three parametrizations the values obtained 
with the raw HFB energies as well as those including the rotational 
energy correction (computed according to our procedure) are given. 
However, the numbers obtained in that way are not very fair in a 
comparison with BCPM's value because in two cases (D1S and D1N) the 
binding energies of a few spherical nuclei were used in the fit, and 
in the other, the binding energy includes quadrupole zero point 
energy corrections not taken into account in our calculations. 
Therefore, in order to provide with a number we can compare with 
BCPM's results we have allowed for a global energy shift of the 
binding energies with shift values obtained as to minimize the 
$\sigma (E)$ values (also given in the table). Both D1S and D1N were 
fitted to the binding energies of just a few spherical nuclei and 
therefore the impact of the rotational energy correction in the 
binding energy of deformed nuclei was not included. The effect of 
adding the rotational correction is noticeable bringing $\sigma (E)$ 
down from its original value of 3.48 (4.88) MeV to 2.15 (2.84) MeV 
for D1S (D1N). The additional phenomenological shift considered 
reduces the raw HFB $\sigma (E)$ value substantially in the two 
cases, but it barely affects the quantity including the rotational 
energy correction in the D1S case and reduces to a surprisingly good 
value of 1.45 MeV the D1N value. On the other hand, D1M is fitted to 
essentially all the experimentally known binding energies and both a 
rotational energy correction and a zero point energy correction 
associated to quadrupole fluctuations was considered for each 
nucleus \cite{D1M}. The raw HFB value given in our table is 
therefore very large but the energy shift reduces substantially the 
$\sigma(E)$ value down to 2 MeV. When the rotational energy is 
included a substantial reduction in $\sigma(E)$ is observed. An 
additional energy shift (meant in this case to mock up the 
quadrupole zero point energy correction, a quantity which is not 
accessible with our computing resources) brings the $\sigma(E)$ 
values down to 1.45 MeV. However, this value is still far from the 
0.7 MeV given in \cite{D1M}. The discrepancy has to be attributed to 
the slightly different rotational energy corrections and the lack of 
quadrupole zero point energies in our calculation. The numbers 
obtained for D1M and D1N are close to the one of BCPM (1.58 MeV) 
when the same set of nuclei and beyond mean field effects are 
considered in all the cases. This fact raises the expectation that a 
BCPM functional fitted with quadrupole zero point energies (and 
eventually taking into account the correlation energy of dynamical 
parity restoration \cite{rob11c} and/or particle number symmetry) 
could lead to a much smaller value of $\sigma(E)$, well below 1 MeV.

Once our model has been established, it is convenient to summarize 
our procedure used in BCPM and to contrast it with the one of the 
more standard techniques employed with Skyrme or Gogny functionals. 
The main difference between the two methods resides in the way the 
bulk part of the functionals is treated. Within BCPM the latter is 
adjusted to results of microscopic G-matrix calculations (using a 
polynomial fit). Only a single parameter is fine tuned within 
$10^{-3}$ which is the energy per particle, E/A, at the minimum 
point. Only the remainder of the functional, which accounts for the 
in-homogeneity of the system, is fitted with two open parameters to 
the binding energies of nuclei. So in total there are three 
parameters which are adjusted to experimental data: E/A, a range 
parameter, and the strength of the spin-orbit term. On the other 
hand, in Skyrme or Gogny functionals, {\it all} parameters, bulk and 
finite size, typically of the order of twelve, are in general 
simultaneously adjusted to experiment, sometimes also to a mix of 
experiment and theoretical data as the neutron equation of state, 
for instance. We think the main advantage of our BCPM functional is 
this decoupling of parameters: those implicitly entering the 
polynomial fit are never used to fine tune the binding energy of 
finite nuclei exception made of $E/A$. 

In Ref \cite{Ber05} a linear refitting procedure has been proposed to 
improve the rms deviation of binding energy differences of various 
Skyrme like EDF. The idea is to start from the parameters of those EDF
and build up gradient and hessian information that is analyzed using 
singular value decomposition (SVD) techniques. From the SVD analysis
it is concluded that there are three linear combinations of parameters
which are relevant: the most important is connected to the volume energy
whereas the next two are linear combinations of surface and symmetry energy
coefficients. By performing a refined fit using only those combinations
a spectacular improvement of $\sigma (E)$ is obtained for SLy4 going
from 3.3 MeV for the 579 even-even nuclei of the 2003 mass table to the
value of 1.7 MeV. This result goes along our findings: once the set
of parameters is close to reasonable values only the volume and surface
terms (and eventually the symmetry energy) are relevant. 

A comparison with Skyrme like EDFs is also possible, but given the large
number of parametrizations available we will focus on some recent parametrizations
quoted in Ref \cite{Erl12}. We first mention the UNEDF0 parametrization \cite{Kor10}
that includes in the experimental data set to fit not only binding energies
but also information on radii and pairing gaps. Special attention is paid also
to reproduce nuclear matter properties at saturation. The fit includes data on both
spherical and deformed even-even nuclei and gives an rms for the binding
energy of 520 even-even nuclei of 1.45 MeV. This is a better value than
the one given by BCPM but it is fairly close to it. As a consequence of 
the poor performance of UNEDF0 in describing fission
barrier properties in the actinides (see \cite{Nik11} for the significance
of fission and superdeformation data), a new parametrization denoted 
UNEDF1 was generated \cite{Kor12}. It includes extra data on fission barrier heights
and fission isomer excitation energies of several actinides. The
rms for the binding energy of 555 even-even nuclei given by the new
parametrization is $\sigma (E)=1.9$ MeV which is worse than the $\sigma(E)$
obtained by UNEDF0. In Ref. \cite{Klu09} a survey of Skyrme like parametrizations
was performed. Several observables were included in the fit and several
sets of parameters with different nuclear matter properties were considered.
This is a completely different strategy from the one adopted here, but for
the matter of comparison let us quote the rms value of 1.6 MeV for the binding energy of
one of the parametrizations obtained, namely SV-min, using 513 experimental
masses of even-even nuclei \cite{Rei11}. 
On the other hand, the most recent mass model produced by the Montreal-Brussels
collaboration HFB-21 produces a $\sigma (E)$ of 0.58 MeV but for all even and
odd A nuclei (2149 masses). This figure is obviously much better than the
one by BCPM and other approaches discussed above but the number of nuclei
considered is larger and also the mass model includes some phenomenological
aspects not considered here. 

\begin{table}
	\begin{tabular}{|c|c|c|c|}\hline
	    $\sigma (E)$       &  D1S &  D1M &  D1N \tabularnewline
\hline \hline
        HFB                & 3.48       & 5.08        & 4.88       \tabularnewline
\hline 
        HFB+E$_{ROT}$      & 2.15       & 2.96        & 2.84       \tabularnewline
\hline 	
        HFB + Shift        & 2.53 (2.4) &  2.02 (4.7) & 2.02 (4.5) \tabularnewline
\hline 
        HFB+E$_{ROT}$+Shift& 2.14 (0.2) &  1.47 (2.6) & 1.45 (2.4) \tabularnewline
\hline 		
	\end{tabular}
\caption
{The $\sigma (E)$ values (MeV) for the three parametrizations of the Gogny force
D1S \cite{D1S}, D1M \cite{D1M} and D1N \cite{D1N} and for different kinds
of theoretical calculations. For the row marked as HFB the theoretical
binding energy corresponds to the raw mean field energy, in the one denoted 
HFB+E$_{ROT}$ the rotational energy correction is also included. The last two
rows correspond to the same theoretical set up as before but this time a 
global shift has been applied to the binding energy differences as to minimize the
$\sigma (E)$ value. The values of the shift parameter (in MeV) for each situation are 
given in parenthesis.
\label{tab:Gogny}
}
\end{table}

\subsection{Role of the low density part in the BCPM functional}

\begin{figure}
\includegraphics[width=\columnwidth]{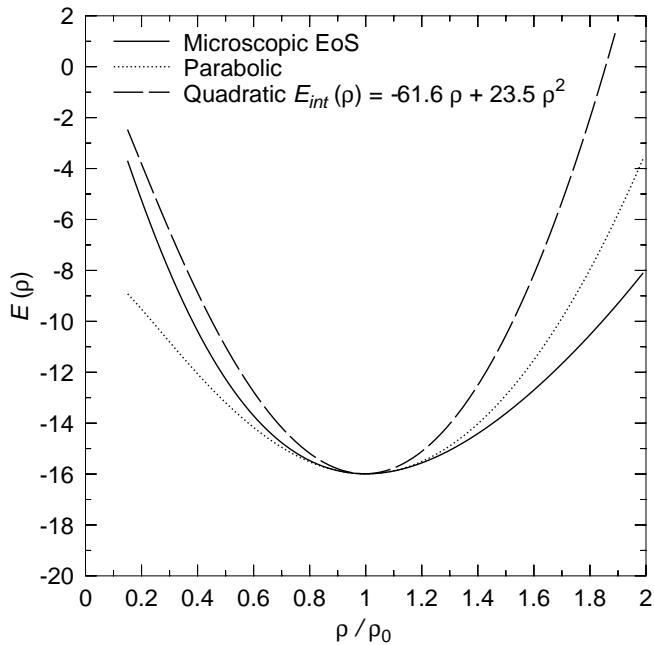}
\caption{The different equations of state considered are plotted as a function
of the density $\rho$ in units of the saturation density, $\rho_{0}$.
The microscopic EOS is the one used to derive the BCPM functional
(full line), whereas the parabolic one (dotted line) is the one obtained
by a parabolic fit to the microscopic EOS at saturation. Finally,
the quadratic EOS is given by the dashed line. In the legend the parameters
of the quadratic EOS are shown.}
\label{Fig5} 
\end{figure}

One may wonder how significant is the detailed density 
behavior of the microscopic input below saturation. For instance the 
low density part gives more attraction than most of the more 
phenomenological mean field approaches. It is not the purpose of 
this paper to investigate this question in detail here as it was already discussed
in an earlier paper \cite{Baldo.04}. However, for a first test of 
the  ingredients that make up our functional, we have analyzed the 
impact of considering instead of the full Equation of State (EOS) a 
parabolic approximation to it. The parabolic approximation is 
obtained from the series expansion of the microscopic EOS around 
saturation $E_{\textrm{Par}}(\rho)\approx 
E(\rho_{0})+k(\rho-\rho_{0})^{2}$. It has the obvious disadvantage 
of reaching a finite value at zero density, which is unphysical. In 
order to correct this unwanted behavior another quadratic 
approximation has also been considered, namely 
$E_{\textrm{Quad}}=\epsilon_{1}\rho+\epsilon_{2}\rho^{2}$. It goes 
to zero at zero density, and the two parameters are adjusted to have 
the correct minimum at saturation. As a consequence of the limited 
number of parameters, it is not possible to modify the curvature of 
$E(\rho)$ around saturation. In Fig. \ref{Fig5} we plot the three 
EOS considered as a function of density. 

\begin{figure}
\includegraphics[angle=-90,width=\columnwidth]{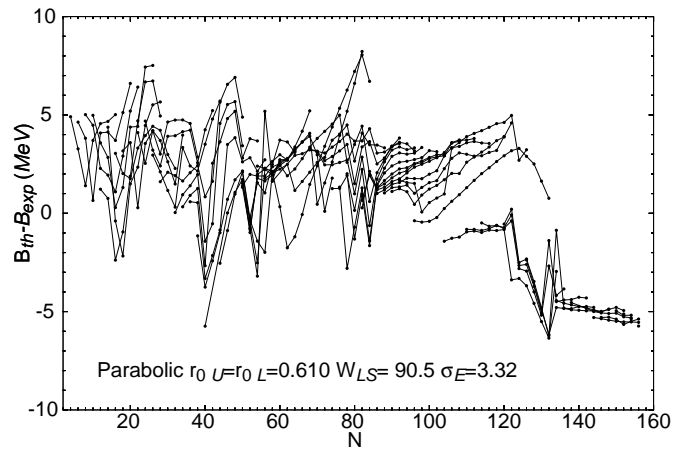}
\caption{Binding energy differences $\Delta B=B_{\textrm{th}}-B_{\textrm{Exp}}$
(MeV) as a function of N obtained for the energy density functional
corresponding to the quadratic approximation 
$E_{\textrm{Quad}}=\epsilon_{1}\rho+\epsilon_{2}\rho^{2}$.}
\label{Fig:EOSquad}
\end{figure}

The finite nuclei results with the parabolic approximation turn out 
to be rather poor, probably as a consequence of the behavior at 
zero density. On the other hand, the quadratic approximation results 
turn out to be quite reasonable in spite of the $\sigma(E)$ value of 
around 10 MeV obtained without fitting the $r_{0L}=r_{0U}$ range 
parameters. This high value comes essentially from super-heavy 
nuclei (see below). If the range parameters are searched for to 
minimize the $\sigma(E)$ we obtain a much reduced value of 3.32 (see 
Fig. \ref{Fig:EOSquad}). The binding energy difference plot still 
shows the problems with the super-heavies mentioned above and we 
observe that the drop in the binding energy difference for neutron 
numbers in excess of 130 is completely different from the rest which 
looks rather reasonable. The behavior of the super-heavies may have 
to do with the fact that there is an extreme balance of Coulomb, 
surface and shell energies and any slight perturbation of the full 
solution may lead to a strong failure. In conclusion, we can say 
that a quadratic fit to the EOS may work reasonably well but for 
optimal numbers, the details of the microscopic density dependence 
of the EOS below saturation are important.

\section{Performance of the new parametrization for other quantities}

Nuclear binding energies play an important role in the description 
of atomic nuclei but there are other relevant observables in 
nuclear dynamics and therefore any functional 
expected to become ``universal'' has to perform well also in the 
description of those quantities. We now discuss a couple of 
the most relevant ones.

\subsection{Radii}

\begin{figure}
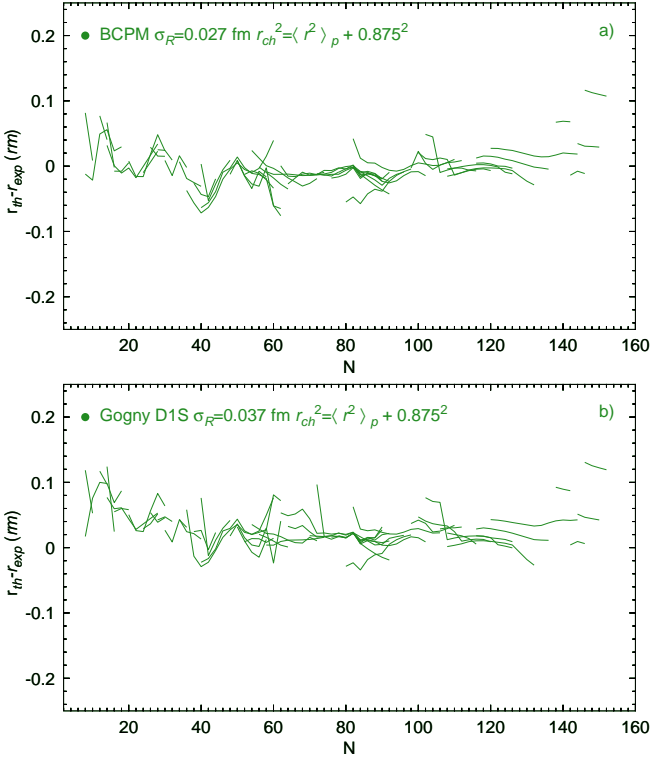

\includegraphics[angle=-90,width=\columnwidth]{DeltaR_15.98_0.659_90.5.ps}

\includegraphics[angle=-90,width=\columnwidth]{DeltaR_D1S}
\caption{Differences $\Delta r=r_\textrm{th}-r_\textrm{exp}$ between the computed radii and
the experimental data taken from \cite{ang04,Buc99,Cam02}. In 
panel a) the results for the BCPM functional and in panel b)
the results for the Gogny D1S force. A proton's radius value of 0.875 fm has
been considered.\label{fig:radii}}
\end{figure}

The nuclear charge radius is a relevant observable connected to many 
other physical quantities. Experimentally, it can be accurately 
measured using electron scattering or other complementary techniques. 
Theoretically, already at the mean field level it is possible to 
obtain reasonable values for charge radii if the proton radius is 
taken into account. The charge radii for all even-even nuclei considered 
has been computed and 
compared to the experimental data published in Angeli's 
compilation \cite{ang04} except for some Sr and Zr isotopes  
where laser spectroscopy results  have been considered instead \cite{Buc99,Cam02}. 
The theoretical predictions (computed with a proton's radius of 0.875 fm \cite{PDG08})
are confronted to the measured values of 313 even-even nuclei in panel a) of 
Fig. \ref{fig:radii}.
A rms value of 0.027 fm is obtained, which is  a quite reasonable number
taking into account that the charge radius has not been 
considered in the fitting protocol. For comparison, the same 
quantity for the same nuclei computed with the Gogny D1S force 
(see panel b of Fig. \ref{fig:radii}) is 
0.037 fm and 0.028 fm with the more recent parametrization D1M \cite{D1M}. 
We notice that the largest contributions to the BCPM rms value 
come from the heaviest nuclei considered, where a strong deviation between theory and 
experiment is observed. Systematic deviations are also observed in  
nuclei around N=40, N=60 and N=80, which are regions of the Nuclide chart 
characterized by the phenomenon of shape coexistence. As for the 
binding energies, the deviations strongly fluctuate in light nuclei 
with $N<40$. The same pattern is also observed for the Gogny D1S 
results, suggesting that the origin for the discrepancies is 
more likely connected with a poor description of the ground state than with 
deficiencies of the interactions. A comparison of the two plots 
reveals that a better figure for $\sigma_{R}$ could be achieved in 
BCPM if an overall displacement would be performed, i.e. the 
theoretical values systematically underestimate the experimental 
values. It may be that size fluctuations (RPA correlations) bring 
the radii to their correct value as suggested in \cite{Dupuis.06}. 
This is a task for the future.

\subsection{Quadrupole and octupole deformations}

\begin{figure}
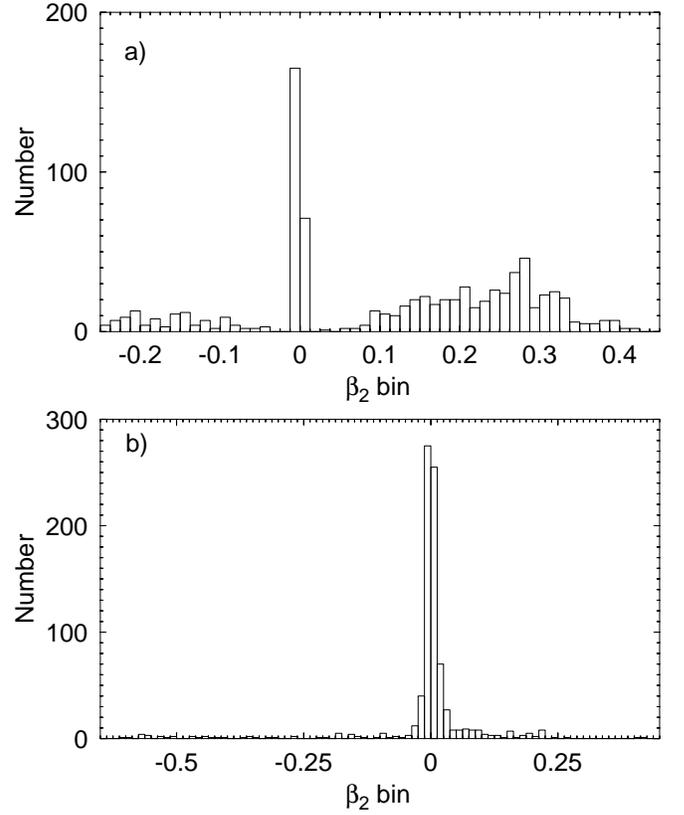

\includegraphics[width=\columnwidth]{HIST1}%

\includegraphics[width=\columnwidth]{HIST}
\caption{The number of nuclei with a given quadrupole
deformation parameter $\beta_{2}$ in their ground state (in bins
0.0125 units wide) are plotted versus $\beta_{2}$ in  panel a). 
In panel b) the number of nuclei with a difference in the ground 
state deformation parameter $\beta_{2}$ 
obtained with BCPM and Gogny D1S are given. The width of the $\beta_{2}$
bins is the same as in panel a).}\label{FigBeta2}
\end{figure}

The quadrupole deformation parameter of the ground state is another 
relevant parameter associated to low energy nuclear properties like 
$2^{+}$ excitation energies or $B(E2)$ transition probabilities. The 
connection of the intrinsic deformation parameter $\beta_{2}$ with 
experimental $B(E2)$ transition probability values is somehow 
uncertain as it relies on the strong deformation limit of angular 
momentum projected theories to obtain the rotational formula \cite
{man75}. For this reason we have preferred to compare BCPM's 
predictions with the ones of a well reputed, performing and 
predictive effective interaction. We have chosen the Gogny force, 
D1S, (results for the most recent published parametrization of the 
Gogny force, D1M, are very similar). In the upper panel of Fig. \ref
{FigBeta2} a histogram is plotted depicting the number of nuclei 
with ground state quadrupole deformation parameters $\beta_2$ 
between $\beta_{2}(n)$ and $\beta_{2}(n)+\delta \beta_{2}$. The 
discrete $\beta_{2}(n)$ values are given by the sequence 
$\beta_{2}(n)=\delta\beta_{2} n$ with $n=\ldots,-2,-1,0,1,2,\ldots$ 
The value $\delta\beta_{2}=0.0125$ has been chosen in such a way 
that each bin represents roughly 5$\%$ of the typical value of 0.25 
for the quadrupole deformation parameter. On the lower panel we plot 
a histogram with the difference 
$\beta_{2}(\textrm{D1S})-\beta_{2}(\textrm{BCPM})$. We observe how 
most of the 818 nuclei considered have a difference of less than 
0.0125 in absolute value. A more detailed analysis of the results 
reveals that most of the discrepancies take place in the region 
$Z\approx N\approx40$ which is widely recognized as a region of 
prolate-oblate shape coexistence and triaxiality.

Octupole deformation is associated to the multipole operator of 
order three and breaks the parity symmetry. It is not as common as 
quadrupole deformation in the ground state of atomic nuclei as very 
specific combinations of orbitals around the Fermi surface have to 
occur to favor it. Known regions showing octupole deformation are 
the region around the $^{224}$Ra (Z=88, N=136) and the region around 
$^{144}$Ba (Z=56 and N=88). We have tested the octupole properties 
of BCPM by relaxing the parity symmetry in our self-consistent 
calculations. In this way, the system can end up in an octupole 
deformed configuration in its quest for the minimum energy. The 
results show energy gains of the order of a few hundred keV and up 
to 1 MeV in nuclei in the aforementioned regions and both the energy 
gains and values of the $\beta_{3}$ deformation parameters are 
compatible with those obtained with the Gogny interaction 
\cite{rob11c}. Overall, the octupole correlation energies are about a few 
hundred keV smaller than the corresponding ones obtained with Gogny 
D1S and D1M ones. Typical examples are the case of $^{224}$Ra with 
octupole correlations energies of 0.63 MeV and 1.31 MeV for BCPM and 
Gogny D1S, respectively; or the one of $^{144}$Ba where the 
corresponding values are 0.15 MeV and 0.68 MeV for BCPM and Gogny 
D1S respectively. For D1S there are five nuclei with mean field 
octupole correlation energies larger than 1 MeV and another nine 
with energies between 1.0 and 0.5 MeV. For BCPM the numbers are two 
with energies above 1 MeV (by just a few keV) and nine in the 
interval 1.0 to 0.5 MeV. The situation is similar to the case of 
BCP1 studied in Ref. \cite{Robledo.10} with the only difference that 
in BCP1 the mean field octupole correlation energy was comparable to 
the one of Gogny D1S. However, as a consequence of the larger 
collective masses, the excitation energies predicted by BCP1 were 
substantially lower than the ones of Gogny D1S and the experiment. 
It remains to be tested what would be the spectroscopic predictions 
of BCPM concerning negative parity excited states but the lower mean 
field correlation energies go in the right direction.

\subsection{Fission properties}

Another testing ground for quadrupole and octupole properties of 
atomic nuclei is the shape evolution from the ground state to 
spontaneous fission. For this reason we have performed fission 
barrier calculations for a few selected examples and the results 
obtained for fission barrier heights and widths as well as mass 
asymmetry near scission (connected with fragment's mass asymmetry) 
are quite close to the results obtained with D1S, a functional that 
was specifically tailored to describe fission properties \cite{D1S}. 
A detailed account of these calculations will be presented elsewhere 
and here we will focus on describing the results for a couple of 
examples: one is the paradigmatic case of $^{240}$Pu whose fission 
properties have been computed with almost any proposed interaction. 
The other is the super-heavy $^{262}$Sg where experimental data on 
spontaneous fission exist. In both cases, the results will be 
compared with the benchmark calculations carried out with the Gogny 
EDF. For an early account of fission barrier properties with 
previous versions of the BCP functional (BCP1 and BCP2) the reader 
is referred to Ref. \cite {rob11b}. 

\begin{figure}
\includegraphics[width=0.5\textwidth]{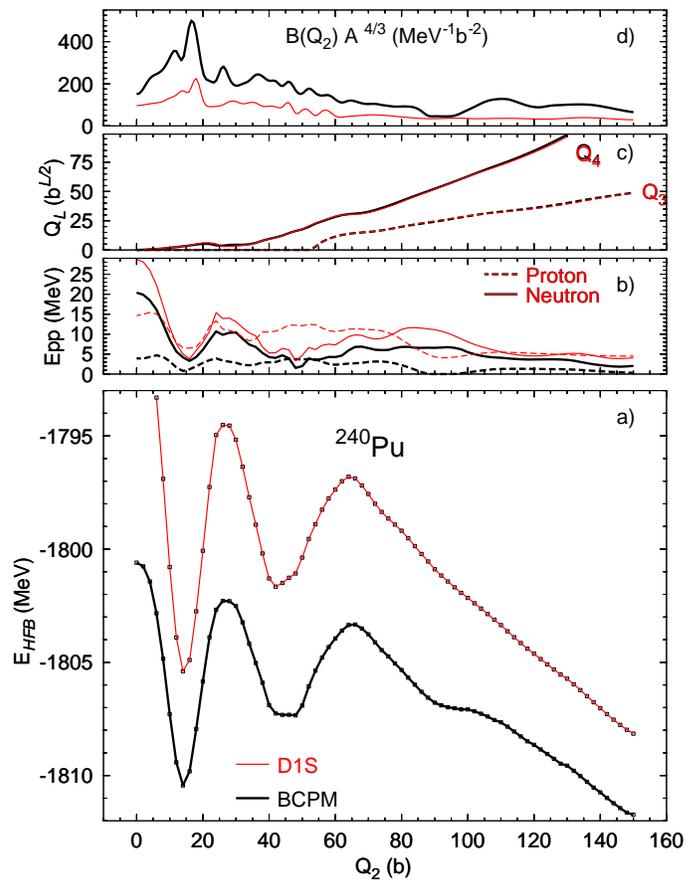}
\caption{(Color online) Fission properties of $^{240}$Pu 
obtained with the Gogny D1S interaction (red) and the BCPM functional (thick black).
In  panel a) the HFB energy is depicted as a function of the mass 
quadrupole moment, from the spherical up to a very elongated 
configuration corresponding to $Q_{2}$=150 b. In the other three panels, quantities relevant for the 
fission half life are given: in panel b) the particle-particle correlation
energies for protons (dashed lines) and neutrons (full lines); in panel c) the octupole
and hexadecapole moments; and finally in panel d) the collective mass for 
quadrupole motion.\label{FISPu}}
\end{figure}

\begin{figure}
\includegraphics[width=0.5\textwidth]{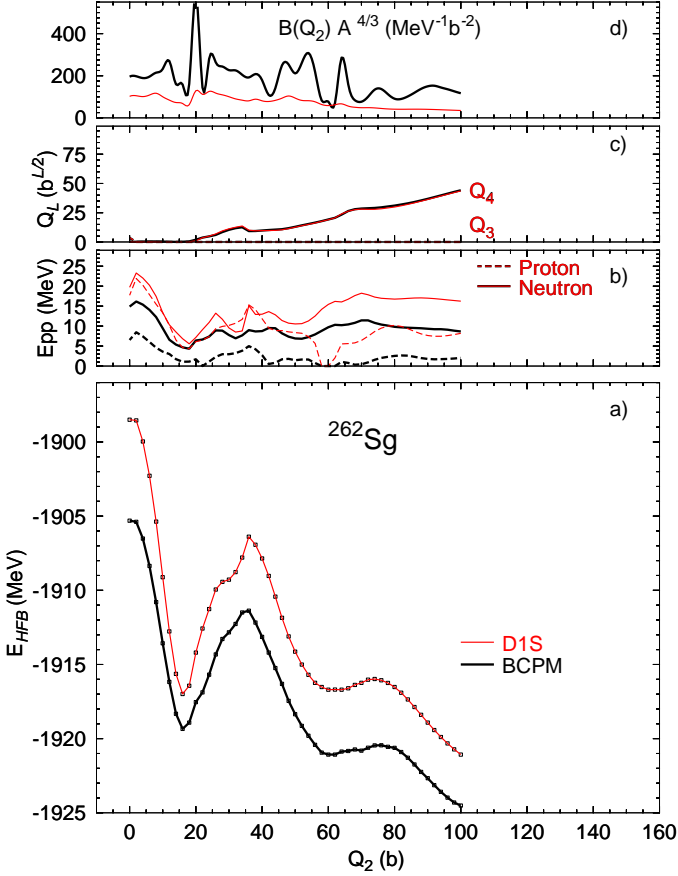}%
\caption{(Color online) Fission properties of the super-heavy nucleus $^{262}$Sg. 
See Fig. \ref{FISPu} for the figure caption.\label{FISSg}}
\end{figure}

We have followed the standard procedure that consists in the 
evaluation of the HFB energy as a function of the mass quadrupole 
moment $Q_{20}$ with the other multipole moments free to adopt any 
value in order to minimize the energy. Along this path to fission, we 
compute the collective inertias $B(Q_{20})$ required to evaluate the 
spontaneous fission half life in the WKB approximation (see 
\cite{rob11b} for details). The relevant quantities are shown in Figs. 
\ref{FISPu} and \ref{FISSg} for the nuclei $^{240}$Pu and $^{262}$Sg, respectively. In the 
lower panels, the HFB energy is depicted as a function of $Q_{20}$ 
for both BCPM and Gogny D1S cases. In the two considered nuclei the 
shape of the energy landscape looks rather similar regardless of the 
interaction considered. The maximum and minimum are located roughly 
at the same position and it is only the height of the barriers which 
changes with the interaction. Overall, Gogny D1S produces barriers 
higher than BCPM in agreement with the larger surface energy coefficient
of D1S. The relevance of this characteristics of BCPM 
depends upon the reduction of the height of the first fission 
barrier when triaxial shapes are included in the calculation. In the 
case of the Gogny forces, the reduction can be as large as two or 
three MeV but we do not have a hint for BCPM yet. Work is in 
progress to explore the characteristics of BCPM regarding triaxial 
shapes. In  panel c) of Figs. \ref{FISPu} and \ref{FISSg}, the octupole 
and hexadecapole moments are depicted as a function of $Q_{20}$ for both 
interactions. The results for Gogny D1S and BCPM lie one on top of each 
other and can not be distinguished in the plot. We observe that in 
$^{240}$Pu octupole deformation develops in its way to fission 
pointing to a dominant mass asymmetric fission mode. On the other 
hand, $^{262}$Sg remains reflection symmetric along the whole 
fission path and therefore symmetric fission is expected to be the 
dominant mode in this nucleus.  In panel b) of both figures the 
particle-particle pairing energies are given both for protons 
(dashed) and neutrons (full) for the two interactions. Again, the 
agreement in the evolution of these quantities with $Q_{20}$ for the 
two interactions considered is very good but BCPM produces smaller 
pp pairing energies suggesting weaker pairing correlations. The 
collective inertias are quantities very sensitive to the amount of 
pairing correlations as they strongly depend on the excitation 
energies of the low lying two quasiparticle excitations. In panel d)  
of Figs. \ref{FISPu} and \ref{FISSg} it is observed that the 
collective inertias of BCPM are between 2-3 times larger than 
the ones of Gogny D1S. The impact of the larger inertias (see \cite
{rob11b} for a thorough discussion of this issue) on the spontaneous 
fission half lives is to make them longer. This effect is the 
opposite of having lower barrier heights in BCPM as compared to D1S. 
Therefore, the final values of the half lives will 
depend on the specific values of the quantities entering the 
"collective action". If the spontaneous fission half lives are 
computed in the WKB approximation we obtain for the $^{240}$Pu case 
the values $t_{1/2} (\textrm{sf}) = 6.2 \times 10^{18} \textrm{s}$ 
for D1S and $t_{1/2} (\textrm{sf}) = 1.2 \times 10^{28} \textrm{s}$ 
for BCPM, to be compared to the experimental value of $3.5\times 
10^{18} \textrm{s}$. For the $^{262}$Sg case we obtain the values 
$t_{1/2} (\textrm{sf}) = 2.96 \textrm{s}$ for D1S and $t_{1/2} 
(\textrm{sf}) = 4.2 \textrm{s}$ for BCPM, to be compared to the 
experimental value of $ 6.9 \textrm{ms}$. From the above results we 
conclude that BCPM tends to yield larger values of $t_{1/2} 
(\textrm{sf})$ than Gogny D1S and those values can be several orders 
of magnitude larger than in the Gogny D1S case. This is a direct consequence of 
the larger collective inertia and therefore a direct consequence of 
the reduced pairing correlations in BCPM. This results could lead us to the 
conclusion that BCPM should include stronger pairing correlations if 
the spontaneous half lives of fission are to be improved. However, 
we have to be aware of the impact on small changes in the parameters 
entering the WKB formula. For instance, the ground state energy 
usually incorporates some sort of zero point energy correction 
(denoted $\epsilon_0$ in the literature) that strongly influences 
the WKB values of $t_{1/2} (\textrm{sf})$. Its value has to be 
connected to the quantum effects associated to quadrupole motion but 
its exact meaning is still uncertain. As an example of the 
sensitivity of $t_{1/2} (\textrm{sf})$ to this parameter, let us 
mention that if we use a value of $\epsilon_0$=2.5 MeV instead of 
the value $\epsilon_0$=1.5 MeV used originally we will obtain half 
lives which are 6 to 12 orders of magnitude smaller. In systematic 
fission calculations \cite{war02} the value of $\epsilon_0$ is fine 
tuned for each isotopic chain considered and therefore absolute 
values for some choice of $\epsilon_0$ as given here should not be 
considered too seriously for a comparison with experiment. Finally, the 
agreement between BCPM and Gogny D1S in the values of $Q_{30}$ along 
the fission path indicates that both interactions will produce, at 
least in a static framework, the same fission fragment mass 
distributions.

There are three parameters that characterize in a quantitative way the
fission process of $^{240}$Pu, one is the height of the first fission 
barrier $E_A$, the second is the excitation energy of the fission isomer
with respect to the ground state $E_{II}$ and finally we have the height
of the second barrier $E_B$. The accepted experimental values for those parameters
are $E_A=6.05$ MeV, $E_{II}=2.8$ MeV and $E_B=5.15$ MeV whereas the BCPM HFB predictions
are $E_A=8.15$ MeV, $E_{II}=3.09$ MeV and $E_B=7.10$ MeV. We notice a rather good
agreement specially taking into account that no fission data has been 
considered in the fit. The BCPM value of $E_A$ is somewhat too high but 
this can be attributed to not having considered triaxiality in the 
calculation, an effect that can lower the energy by a couple of MeV. If the
rotational energy correction is considered, the
theoretical predictions get reduced to 7.04, 1.69 and 5.31 MeV respectively.
These values are much closer to the experimental value than the HFB ones.
The only value that seems to fall too low is $E_{II}$ but the trend is
consistent with recent estimations that reduce the experimental value
to 2.25 MeV in $^{240}$Pu \cite{Hun01}.

\section{Excitation properties of the BCP energy density functional}

Although the BCPM energy density functional is basically built up to 
deal with some properties of the ground state such as energies and 
radii, it is however interesting to check if this functional can also 
provide useful information about some excited nuclear states, as for 
example the isoscalar giant monopole and quadrupole resonances, 
ISGMR and ISGQR, respectively. Giant resonances can be understood as 
small amplitude oscillations of nuclei that are the response to an 
external field generated by hadronic or electromagnetic probes. A 
useful theoretical framework for describing these oscillations is 
the RPA \cite{RS80} that allows to obtain the particle-hole strength 
function $S(E)$ which describes the response of the nuclei. However, 
if the strength is concentrated in a narrow region of the energy 
spectrum, as it is for example the case of the monopole and 
quadrupole oscillations in medium and heavy  stable nuclei, the so 
called sum rule approach \cite{boh79} is a useful tool that allows 
to estimate the energy of the giant resonances without performing 
the full RPA calculations. The sum rules are defined as the moments 
of the strength function as
\beq
m_k = \int_0^{\infty} E^k S(E) E
\label{000} 
\eeq

The calculation of the sum rules is only easy to handle in few 
particular cases. Examples are just the ISGMR and ISGQR ones. In 
these cases the sum rules needed to estimate the average excitation 
energy of these resonances can be simplified. Of particular interest 
are the cubic-energy, energy and inverse-energy 
weighted sum rules, $m_3$, $m_1$ and $m_{-1}$, respectively. Once 
they are determined one can obtain two estimates of the excitation 
energy of the ISGMR and ISGQR as
\beq
\bar{E}_3 = \sqrt{\frac{m_{3}}{m_{1}}}
\quad
\textrm{and}
\quad
\bar{E}_1 = \sqrt{\frac{m_{1}}{m_{-1}}}.
\label{eq00} 
\eeq 

Details about the scaling method and constrained HF calculations, which allow to 
determine the $m_3$ and $m_{-1}$ sum rules, respectively, is given 
in the Appendix. In the following we will refer to the average 
energies provided by the scaling method $\bar{E}_3$ in Eq.  (\ref{eq19a}) 
and constrained HF calculations $\bar{E}_1$ in Eq. (\ref{eq19}) as scaled 
and constrained energies, respectively. 


\begin{table}
\begin{center}
\caption{\label{T1} Theoretical $E_3$ and $E_1$ estimates of the average 
excitation energy of the GMR including pairing correlations. The $E_3$ 
estimate of the GQR, also including pairing, is also displayed. The 
experimental energy of the centroid and the corresponding error for the 
GMR and GQR are also given.}
\vspace{0.5cm}

\begin{tabular}{|c|c|c|c|c|c|}
\hline \hline
 Nucleus & $E_3(M)$ & $E_1(M)$ & $E_3(Q)$ & Exp(M)& Exp(Q) \\ \hline
$^{90}$Zr  & 19.06 & 18.32 & 13.34 & 17.81 $\pm$ 0.32 & 14.30 $\pm$ 0.40  \\
$^{144}$Sm & 16.44 & 15.62 & 11.45 & 15.40 $\pm$ 0.30 & 12.78 $\pm$ 0.30 \\
$^{208}$Pb & 14.49 & 13.84 & 10.16 & 13.96 $\pm$ 0.20 & 10.89 $\pm$ 0.30 \\
$^{112}$Sn & 17.75 & 16.83 & 12.36 & 16.1 $\pm$ 0.1 & 13.4 $\pm$ 0.1 \\
$^{114}$Sn & 17.64 & 16.75 & 12.28 & 15.9 $\pm$ 0.1 & 13.2 $\pm$ 0.1 \\
$^{116}$Sn & 17.53 & 16.66 & 12.21 & 15.8 $\pm$ 0.1 & 13.1 $\pm$ 0.1 \\
$^{118}$Sn & 17.41 & 16.55 & 12.15 & 15.6 $\pm$ 0.1 & 13.1 $\pm$ 0.1 \\
$^{120}$Sn & 17.29 & 16.43 & 12.09 & 15.4 $\pm$ 0.2 & 12.9 $\pm$ 0.1 \\
$^{122}$Sn & 17.18 & 16.32 & 12.04 & 15.0 $\pm$ 0.2 & 12.8 $\pm$ 0.1 \\
$^{124}$Sn & 17.06 & 16.21 & 12.44 & 14.8 $\pm$ 0.2 & 12.6 $\pm$ 0.1 \\
$^{106}$Cd & 18.09 & 17.07 & 12.70 & 16.50 $\pm$ 0.19 &     \\ 
$^{110}$Cd & 17.85 & 16.97 & 12.49 & 16.09 $\pm$ 0.15 & 13.13 $\pm$ 0.66 \\    
$^{112}$Cd & 17.74 & 16.83 & 12.38 & 15.72 $\pm$ 0.10 &     \\
$^{114}$Cd & 17.59 & 16.73 & 12.29 & 15.59 $\pm$ 0.20 &     \\
$^{116}$Cd & 17.44 & 16.52 & 12.19 & 15.40 $\pm$ 0.12 & 12.50 $\pm$ 0.66 \\  
\hline
\end{tabular}
\end{center}
\end{table}
\begin{table}
\begin{center}
\caption{\label{T2} $E_3$ and $E_1$ estimates of the average
excitation energy of the GMR without including pairing correlations.}
\vspace{0.5cm}

\begin{tabular}{|c|c|c|}
\hline \hline
 Nucleus & $E_3(M)$ & $E_1(M)$ \\ \hline
$^{90}$Zr  & 19.07 & 18.36 \\ 
$^{144}$Sm & 16.46 & 15.77 \\
$^{208}$Pb & 14.49 & 13.84 \\
$^{112}$Sn & 17.81 & 17.07 \\
$^{114}$Sn & 17.64 & 17.71 \\
$^{116}$Sn & 17.56 & 16.66 \\
$^{118}$Sn & 17.42 & 16.55 \\
$^{120}$Sn & 17.30 & 16.44 \\
$^{122}$Sn & 17.19 & 16.33 \\
$^{124}$Sn & 17.08 & 16.21 \\
$^{106}$Cd & 18.10 & 17.36 \\ 
$^{110}$Cd & 17.86 & 17.11 \\    
$^{112}$Cd & 17.74 & 16.98 \\
$^{114}$Cd & 17.59 & 16.73 \\
$^{116}$Cd & 17.44 & 16.52 \\  
\hline
\end{tabular}
\end{center}
\end{table}

\begin{figure}
\includegraphics[width=0.85\columnwidth,angle=-90]{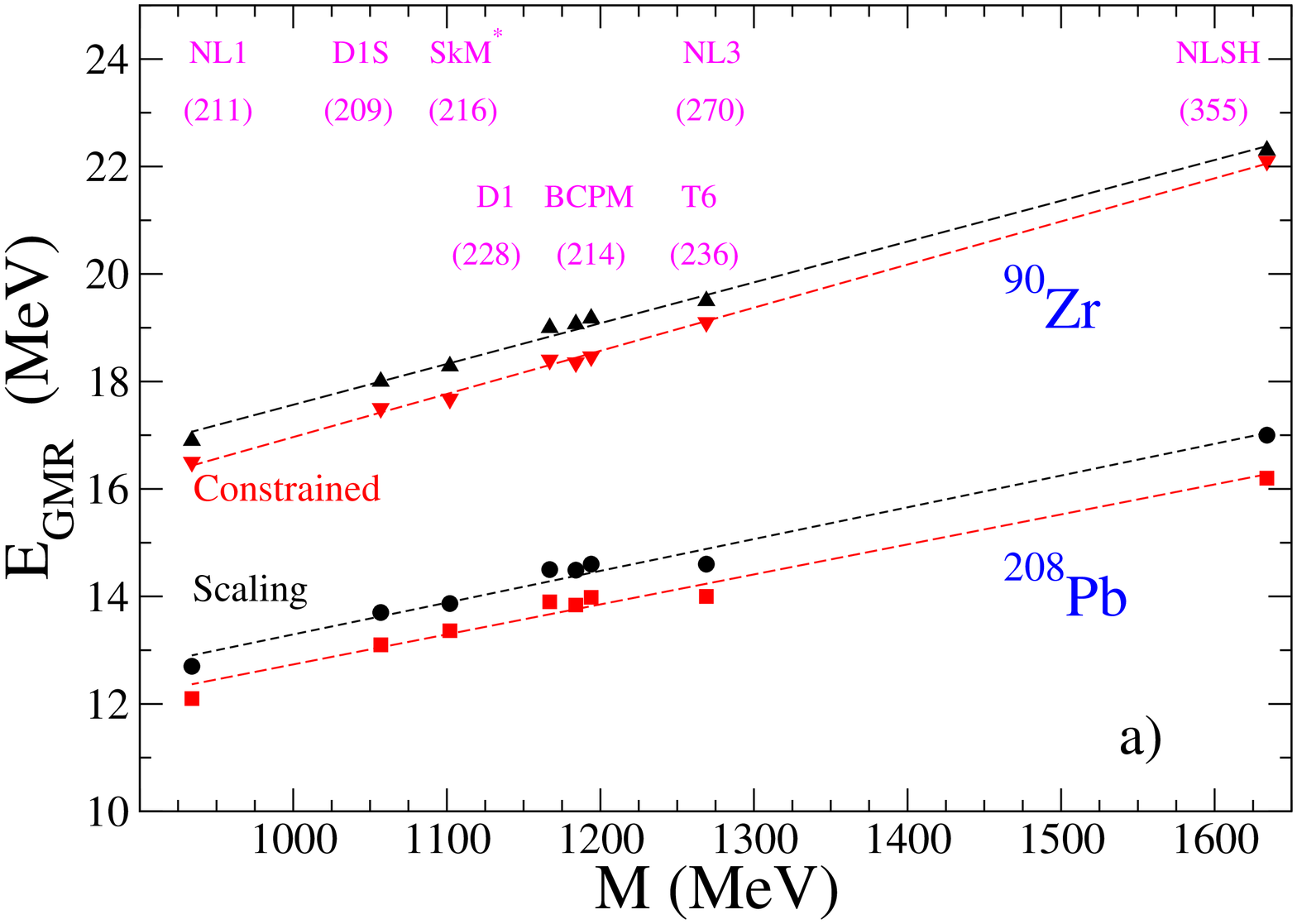}

\includegraphics[width=0.85\columnwidth,angle=-90]{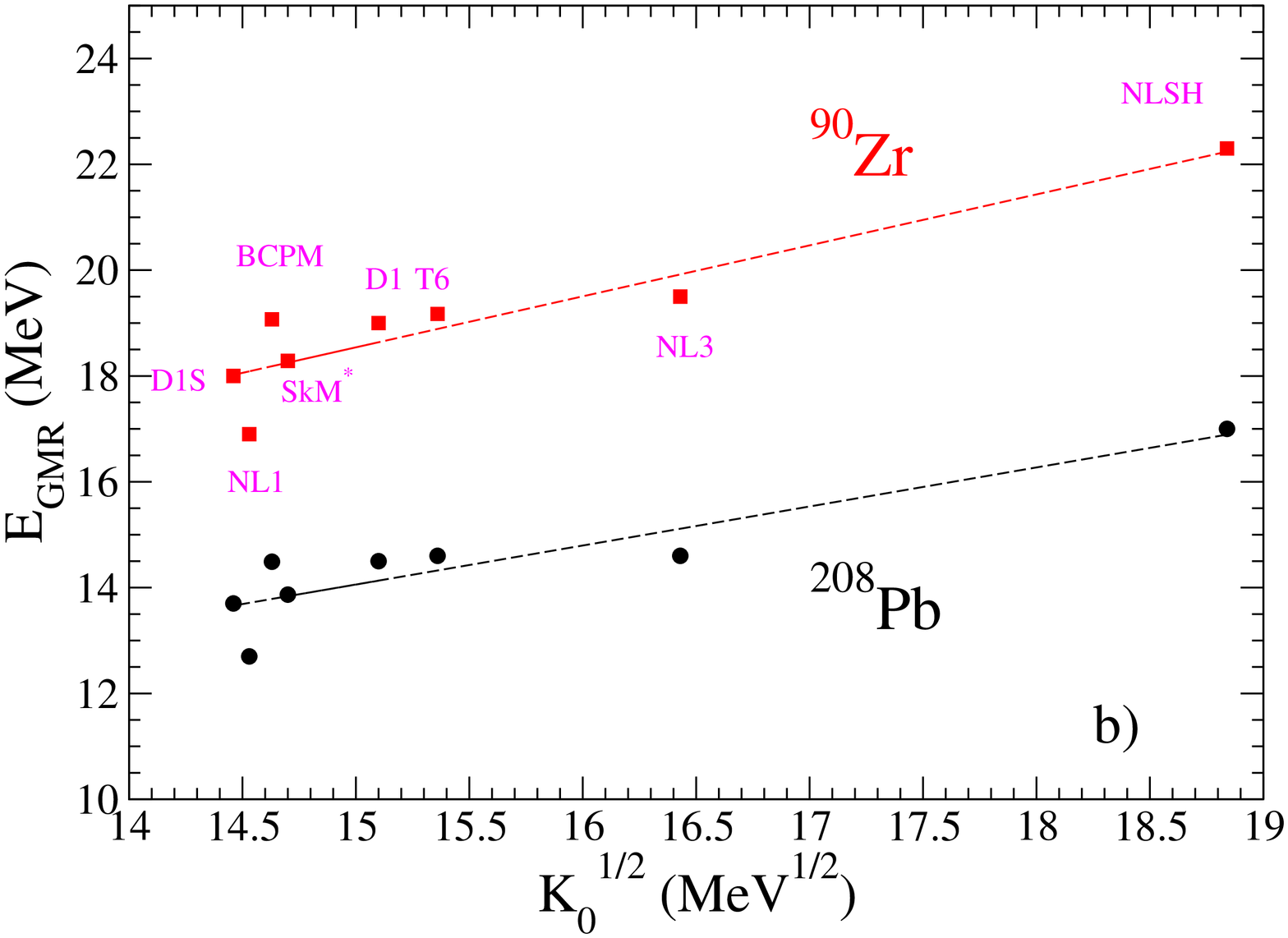}
\caption{\label{F2} (Color online) 
Excitation energies of the GMR of $^{208}$Pb and $^{90}$Zr estimated 
with the scaling method and HF constrained calculations for 
different mean field models as a function of the parameter $M$ 
defined in the text (upper panel). Excitation energies of the GMR of 
$^{208}$Pb and $^{90}$Zr estimated with the scaling approach as a 
function of the square root of the incompressibility modulus (lower 
panel).}

\end{figure}
%
\begin{figure}
\includegraphics[width=0.85\columnwidth,angle=-90]{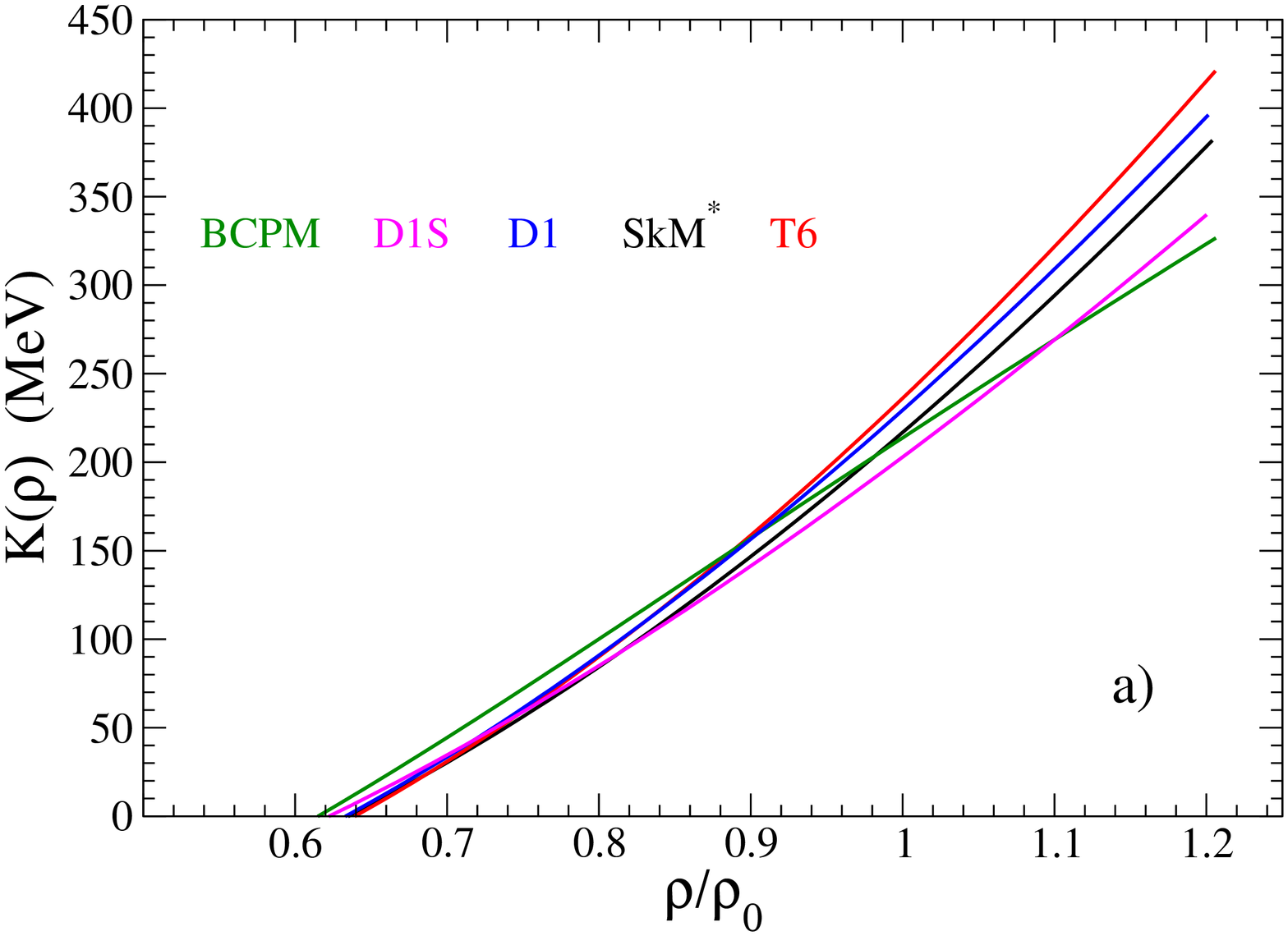}

\includegraphics[width=0.85\columnwidth,angle=-90]{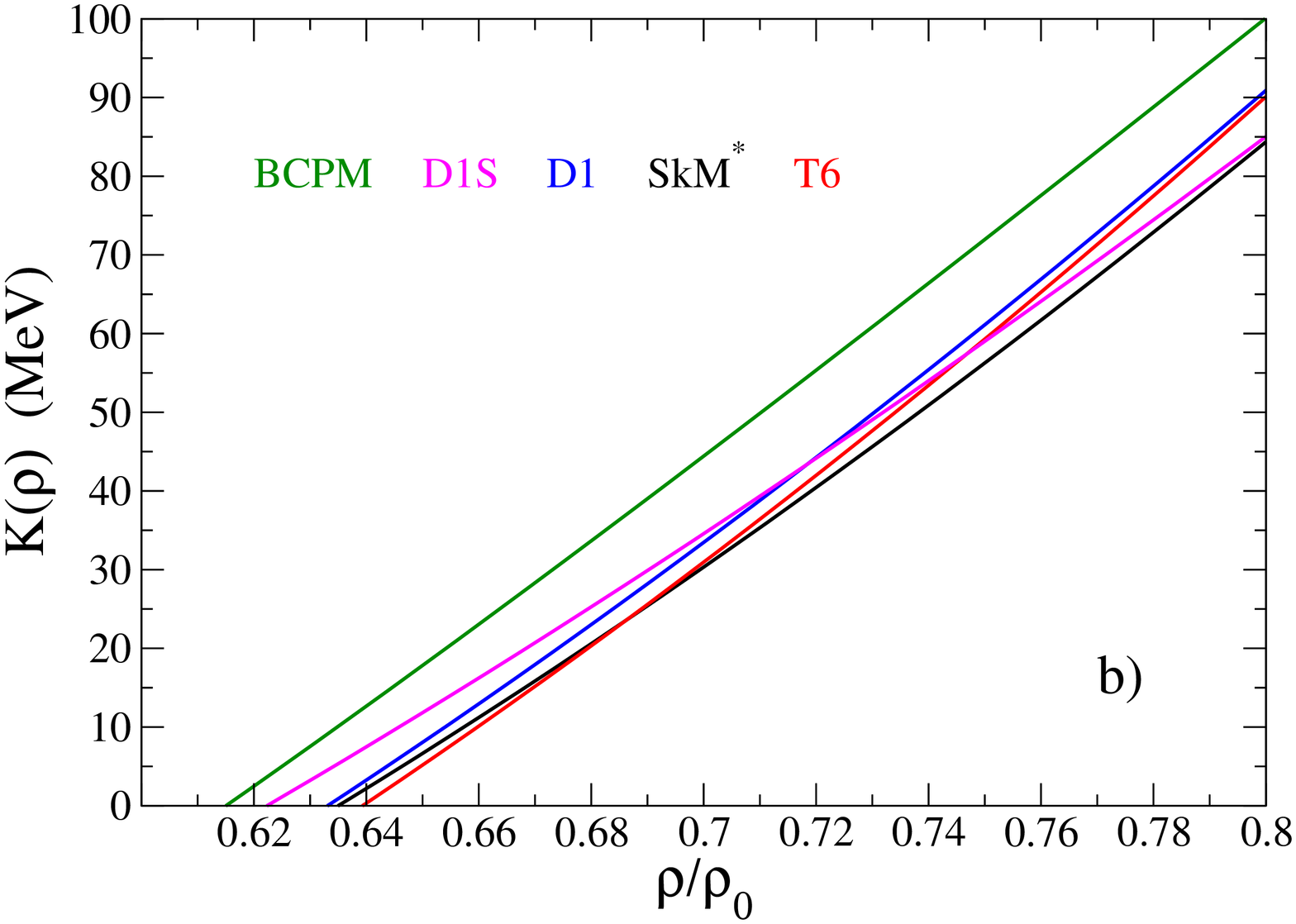}
\caption{\label{F3} (Color online) 
Density dependent incompressibility calculated using different mean 
field models as a function of the density in the ranges 0.5 $\le 
\rho/\rho_0 \le$ 1.25 (upper panel) and 0.6 $\le \rho/\rho_0 \le$ 
0.8 (lower panel).} 
\end{figure}

Experimental information about the excitation energy of isoscalar 
giant resonances in medium and heavy nuclei is obtained through 
inelastic scattering of $\alpha$ particles of few hundreds of MeV 
measured at extremely forward angles. The multipole decomposition 
analysis allows to extract from the scattering data the strength 
associated to different multipolarities, in particular those 
corresponding to monopole and quadrupole. From these strengths, the 
position and width of the ISGMR and ISGQR peaks can be determined 
\cite{you04a,you04b,li07,pat12}.

We analyze first the excitation energies of ISGMR ($E_{GMR}$) of 
several nuclei for which the experimental values are known 
\cite{you04a,you04b,li07,pat12}. We compare these values with the 
theoretical predictions $\bar{E}_3$ and $\bar{E}_1$ provided by the 
BCPM energy density functional. The relevant information about these 
excitations energies is collected in Tables \ref{T1} and \ref{T2}. 
As far as we are including in our study tin and cadmium (neglecting 
the effects of the small deformation) isotopes and $^{144}$Sm, for 
which pairing correlations are essential to describe accurately the 
ground state, we discuss first the role of pairing in the sum rule 
approach used to estimate $E_{GMR}$.

The generalization of RPA to the case of nuclei with open shells 
when pairing correlations are active, is the quasi-particle RPA 
(QRPA). The so-called dielectric theorem, which justifies 
constrained calculations within QRPA to obtain the $m_{-1}$ sum 
rule, has been proven recently by Col\`o and collaborators in \cite
{cap09}. Also Khan et al have shown the validity of the Thouless 
theorem for the energy weighted sum rule in the case of 
self-consistent QRPA based on HFB \cite{kha02}. As far as a simple 
generalization of the Thouless theorem allows to obtain the $m_3$ 
sum rule \cite{boh79}, it can be expected that this also will be the 
case in the QRPA framework. In addition we have also numerically 
checked that the virial theorem, i.e. the stability against scale 
transformations of the wave-functions in Eq. (\ref{eq2}), is also fulfilled 
in the HFB approximation. From this discussion we may conclude that 
the sum rule approach can also be applied in the QRPA case.

In Table \ref{T1} we display the theoretical scaled and constrained 
estimates (including pairing) of $E_{GMR}$ of the nuclei $^{144}$Sm, 
$^{208}$Pb, $^{90}$Zr, $^{112-124}$Sn, and $^{106,110-116}$Cd 
together with the corresponding experimental values \cite 
{you04a,you04b,li07,pat12}. These estimates are given by $\bar{E}_3$ 
and $\bar{E}_1$, respectively, defined previously, which, as 
mentioned, fulfill $\bar{E}_3 > \bar{E}_1$ (see Appendix). The 
escape width (\ref{eq21}) lies in a window of 2-3 MeV, pointing out 
that the monopole strength of the considered nuclei is mainly 
concentrated around well defined peaks. The influence of pairing 
correlations on $E_{GMR}$ has been discussed in earlier literature 
(see \cite {Khan.10,Tselyaev.09} and references therein). We have 
repeated the scaled and constrained estimates of the $E_{GMR}$ 
switching off pairing and the corresponding results are reported in 
Table \ref{T2}. Comparing Tables \ref{T1} and \ref{T2} we notice 
that, when pairing correlations are taken into account (Table \ref
{T1}), the $\bar{E}_1$ and $\bar{E}_3$ values in open shell nuclei 
slightly decrease as compared with the values obtained without 
pairing (Table \ref{T2}) which in agreement with the results of 
Refs. \cite {Khan.10,Tselyaev.09}. The effect of pairing 
correlations is, in general, more important in the constrained 
calculations than in the scaled ones, especially in half-filled 
major shell nuclei as it is the case of $^{112}$Sn, $^{114}$Sn, 
$^{116}$Sn, $^{106}$Cd, $^{110}$Cd, $^{112}$Cd. When the occupancy 
of the shell increases, as it happens in $^{118}$Sn, $^{120}$Sn, 
$^{122}$Sn, $^{124}$Sn, $^{114}$Cd and $^{116}$Cd, the influence of 
pairing is actually very small as it can be seen in Table \ref{T2}.

In order to compare with the experiment, we will use as 
representative the constrained energy $\bar{E}_1$, which estimates 
the energy of the centroid in a very precise way at least in the 
case of the nucleus in $^{208}$Pb \cite{col04,khan09} (see also 
Table IV of \cite{li07} in this respect). From Table \ref{T1}, we 
see that the experimental $E_{GMR}$ in $^{208}$Pb is very well 
reproduced by $\bar{E}_1$ obtained using the BCPM energy density 
functional. The experimental excitation energies of $^{90}$Zr and  
$^{144}$Sm are slightly overestimated by the theoretical constrained 
calculation. However, our theoretical calculation are less 
successful in the case of the cadmium and tin isotopes.  The 
predicted $\bar{E}_1$ values are shifted by around 1 MeV up as 
compared with the experimental results \cite{li07,pat12}. This is a 
general tendency shown by mean-field models of different nature \cite
{col04a,vre03}. Several attempts to explain this behavior have been 
proposed in the past such as the density dependence of the 
incompressibility in neutron rich matter \cite{Piekarewicz.09} or 
different types of pairing interactions \cite{Khan.10}. However, no  
clear explanation has been found to date and the question why Sn 
(and Cd) isotopes are so soft is still an open 
problem. It may be that nonlinear effects coming from particle 
vibration coupling play a role in such nuclei as Sn isotopes, since 
we reproduce the stiff doubly magic nucleus of $^{208}$Pb and not 
the semi-magic ones as, e.g., the Sn isotopes.

From the previous discussion it seems that the $E_{GMR}$ estimated 
with the BCPM functional are somewhat high as compared with the 
experiment in spite of its relatively low incompressibility modulus 
$K_0$=216 MeV (see Table \ref{nmatter}). The fact that effective 
mean field models with different values of $K_0$ can give similar 
values of $E_{GMR}$ in $^{208}$Pb has actually been a general puzzle 
in the past (see \cite{col04a,col04} and references therein). The 
reason is the proportionality between $E_{GMR}$ and the square root 
of $K_{0}$ found by Blaizot \cite{bla95} for Gogny forces. For 
example RMF approaches systematically lead to $E_{GMR}$ comparable 
to the ones obtained with non-relativistic theories like Skyrme or 
Gogny, in spite of the fact that $K_{0}$ values are appreciably 
larger in relativistic than in non relativistic approaches. Some 
time ago Col\`o et al. \cite{col04} showed that one can construct 
Skyrme functionals that give reasonable values for $E_{GMR}$ but 
with relatively high values of $K_{0}$ i.e. values which approach 
the ones of RMF theories. As discussed in \cite{col04}, the 
underlying reason of this fact can be attributed to the different 
density dependence of the symmetry energy at, and around, 
saturation. In the same reference \cite{col04} it is concluded that 
within non-relativistic models there is not a unique relation 
between the value of $K_{0}$ associated to a given model and the 
energy and the $E_{GMR}$ predicted by the same model. 

To clarify this latest problem a very recent proposal of Khan, 
Margueron and Vida\~na \cite{KMV12} may be useful. These authors 
analyzed a relatively large set of Skyrme, Gogny and RMF 
interactions finding that the corresponding density dependent 
incompressibilities $K(\rho)$ defined in Eq. (\ref{incomp}) show a crossing 
point at a density of about 75\% of the saturation density. This crossing density 
corresponds roughly to the average density of a heavy nucleus 
defined as
\beq
\rho_{av} = \frac{1}{A} \int \rho^2(\vec{r}) d \vec{r}. 
\eeq

The authors claim that the excitation energy of finite nuclei is 
more properly correlated with the slope of incompressibility 
$K(\rho)$ computed at the crossing density than with the 
incompressibility modulus computed at saturation density $K_0$. This 
scenario is similar to the one found for the nuclear symmetry energy 
in bulk matter, which also shows a crossing point for a similar 
value of the density ($\simeq$0.11 fm$^{-3}$) \cite{bro00}. In this 
case it is also found \cite{bro00,cen09} that the slope of the 
symmetry energy at the crossing density is linearly correlated with 
the neutron skin of a heavy nucleus, e.g. $^{208}$Pb. In general, 
the existence of a crossing point is actually related to the fact 
that the parameters of successful mean- field models have been 
fitted to reproduce properties of finite nuclei whose densities are, 
on the average, smaller than the saturation density. Therefore, if 
an observable measured in finite nuclei is related to some property 
of nuclear matter, the correlation is better described at the average 
(crossing) density than at the saturation one \cite{KMV12}. 

Following the suggestion of Ref. \cite{KMV12}, we have analyzed the 
behavior of $E_{GMR}$ of $^{208}$Pb and $^{90}$Zr calculated using 
different mean field models as a function of the parameter $M$ 
defined as 
\beq M = 3 \rho \frac{d K(\rho)}{d \rho} \vert_{\rho_
{av}}, 
\eeq 
which is directly related to the slope of the density 
dependent incompressibility $K(\rho)$ at the crossing density 
$\rho_{av}$. In our study in addition of the BCPM energy density 
functional, we have also considered the D1 and D1S Gogny forces, the 
SkM$^*$ and T6 Skyrme interactions and the RMF parametrizations NL1, 
NL3 and NLSH. Using these models the correlation between the scaled 
and constrained estimates of $E_{GMR}$'s and the parameter $M$ are 
displayed in Fig. \ref{F2} for the two aforementioned nuclei. The 
data of the  $E_{GMR}$ predicted by Gogny forces and RMF 
parametrizations are taken from Refs. \cite{sou04} and \cite{pat02} 
respectively.

From this figure it can be seen that the $E_{GMR}$ follows  a linear 
trend with the parameter $M$ (the correlation parameter is r=0.985 
in all the considered cases) with high accuracy, in agreement with 
the suggestion of \cite{KMV12}. Consequently, and as it is claimed 
in \cite{KMV12}, the density dependence of $K(\rho)$ can be 
constrained by the $E_{GMR}$ measured in finite nuclei. The same 
correlation is also fulfilled by the nucleus $^{90}$Zr with $M$ 
evaluated at the same average density. This fact  points out that 
the dependence on the mass number of the average density is actually 
rather weak for medium and heavy nuclei \cite{KMV12}. From the lower 
panel of this figure we also see explicitly  that considering 
globally mean field models of different nature, the linearity 
between $E_{GMR}$ and the square root of the incompressibility 
modulus at saturation \cite{bla95} is lost in some cases, in 
agreement with the conclusion of Ref. \cite{col04}.

To investigate this aspect in more detail, we plot in the upper panel of Fig. 
\ref{F3} the bulk incompressibility of Eq. (\ref{incomp}) as a function of the 
density. In the lower panel we show, magnified, 
the relevant region of the upper panel. For densities close to 0.72$\rho_0$
the density dependent 
incompressibility $K(\rho)$ computed with Skyrme and Gogny forces 
and the RMF takes similar 
values around 40 MeV. However, the BCPM functional predict 
relatively larger values of $K(\rho)$ at this average density. The 
reason of that lies, probably, in the fact that $K(\rho)$ grows more 
linearly with the density than the other considered models. As far 
as the slope of $K(\rho)$ has a value consistent with $E_{GMR}$, as 
it can be seen in Fig. \ref{F2}, the value of $K(\rho)$ at 0.72
$\rho_0$ is larger than the value predicted by the other 
considered mean field models. From the analysis of Fig. \ref{F3} 
one derives two important conclusions. First, the crossing density 
actually disappears when one considers mean field models of 
different nature. What is actually relevant, is to compute the slope 
of $K(\rho)$ at {\it the average density}. The same happens in the 
analysis of the symmetry energy using different mean field models 
\cite{cen09}. Second, the small differences of $K(\rho)$ at the 
average density may explain why $\sqrt{K_0}$ and $E_{GMR}$ are not 
always linearly correlated. This happens, e.g.,  with the NL1 
parametrization with $K_{\infty}$=211 MeV, a value similar to the 
one of the SkM$^*$ force but predicting $E_{GMR}$ clearly smaller, or 
the case of BCPM discussed previously.

In what concerns the ISGQR, we are well aware of the fact that a 
theory with $m*=m$, as is the case of the BCPM functional, 
underestimates the collective quadrupole excitation energy. In Table 
\ref{T1} we display the scaled energies of the quadrupole excitation 
computed with BCPM for several nuclei together with the  available 
experimental values \cite{you04a,you04b,you4c,li07}. Insofar as the 
focus with a KS-DFT approach is on ground state properties, we may 
not worry about the failure in the estimate of the ISGQR. In 
general, for the description of excitation energies, the KS-DFT 
should be generalized to include non-localities. Attempts in this 
direction exist in condensed matter. It may be worth to look into 
that in the future also in the nuclear context.

\section{Conclusions and discussion}

In this work we further elaborated on the previously established 
Kohn-Sham DFT functional, BCP \cite{Baldo.08}, based on advanced 
microscopic nuclear and neutron matter calculations. The BCP 
contained by large 5 adjustable parameters, three for finite size, 
one for spin-orbit and one for bulk. We reduced this number by 
approximately a factor of two coming up with a KS-DFT functional 
containing only three adjustable parameters without deteriorating 
the excellent results. One of these parameters, the strength of the 
spin-orbit force, has the usual value, the results being quite 
insensitive to its variation within a relatively large margin. We 
would like to call this a rather weak, i.e. almost predetermined 
parameter (in addition it may be possible to extract it from the 
G-matrix \cite{Brieva,Hol10,Koh12}). The other two parameters, on the contrary, 
are fixed within extremely tight margins of the order of 10$^{-3}$. 
They concern the bulk energy per particle, E/A, on the one hand and 
the surface energy on the other hand. Indeed in the polynomial fit 
to the microscopically determined EOS, a single fine tuning 
parameter had to be introduced in order to get a minimal rms value 
for nuclear masses as well as a drift free (i.e. scatter around 
zero) difference of theoretical and experimental binding energies as 
a function of mass number. This parameter turned out to pin down the 
binding energy to the value, E/A = 15.98 MeV. We emphasize that the 
fourth digit is significant here. For the second parameter, we made 
the simplifying ansatz to introduce a finite range Gaussian 
convoluting the two densities in the $\rho^{2}$ terms of nuclear and 
neutron matter fits, that is $\rho^{2}\rightarrow\int 
d^{3}r_{1}d^{3}r_{2}\rho({\bf r}_{1})e^{({\bf r}_{1}-{\bf 
r}_{2})^{2}/r_{0}^{2}}\rho({\bf r}_{2})$. The pre-factor is 
determined from the bulk, so that the only free parameter is the 
range $r_{0}$ which evidently is responsible for the surface energy. 
The minimization of the rms of masses (see main text for details) 
again is determined within the very narrow margin of 10$^{-3}$ and 
turned out to have the value $r_{0}=0.659$ fm. With this, a rms 
value for the masses of 1.58 MeV for 579 nuclei and of 0.027 fm for 
the radii with 313 nuclei is  obtained. Those values are as good as, 
e.g., the ones of the Gogny D1S and D1M forces for the same set of 
nuclei. This very encouraging result calls for some comments. It is, 
indeed, quite surprising but on the other hand very satisfying, that 
the two basic nuclear quantities, energy per particle of the bulk 
and surface energy essentially suffice to determine a KS-DFT 
functional with excellent ground state properties of nuclei. We 
should point out that the KS-DFT approach is tailored for the ground 
state and that a time dependent extension is not evident. We 
nevertheless tried to determine at least the GMR energies with the 
sum rule technique and found that the results are in the same ball 
park as the ones of other Skyrme, relativistic, or Gogny type 
approaches. On the other hand the energies of the GQR seem to be 
underestimated systematically by about 1 MeV. This is not so 
surprising, since a particularity of the KS-DFT approach is that the 
bare mass is used while it is known that the GQR is sensitive to the 
value of the effective mass.

Future developments concern eventually the introduction of an 
effective density dependent mass $m^{*} \ne m$. Whether this is 
possible with results as satisfying as here, remains to be seen.
A further point which needs to be improved is pairing. 
At the moment we treat it at a rather elementary level with a zero 
range force and cut off. A finite range force of the Gogny type in 
the pairing channel would certainly be preferable. Promising 
developments in this direction have recently been published \cite{Duke.11}.
At the end let us mention that our theory fulfills all 
the eleven criteria put forward in \cite{stone} besides one. Our 
functional does not agree with the constraint concerning the value 
of the volume isospin incompressibility coefficient that appears in 
the leptodermous expansion of the finite nucleus incompressibility 
\cite{Blaizot.80}. This quantity is defined as 
\begin{equation} 
K_{\tau,v} = \big( K_{sym} -6L + \frac{K'}{K_0}L \big). \label{ktau} 
\end{equation} 
The BCPM functional predicts $K_{\tau,v}$=-195.2 MeV 
which is outside of the range proposed in \cite{stone}, -760 $\le 
K_{\tau,v} \le$ -372 MeV. However, two comments are in order. First 
the extraction of this coefficient from the experimental excitation 
energies of the GMR in finite nuclei \cite{li07,stone,pat12} is not a 
simple task at all and may give erroneous estimates of the 
coefficients of the leptodermous expansion of the finite nucleus 
incompressibility as it has been discussed in the past (see for 
example Refs. \cite{bla95,Patra.02}). Second, there are other 
estimates of the  $K_{\tau,v}$ coefficients available in the literature,
as for instance in Ref. \cite{Chen.09}, that predict -370 $\pm$ 120 
MeV which embraces the prediction of the BCPM functional. In 
general, the criteria and their limitations in \cite{stone} can 
certainly be subjected to debate. For example there is no reason to 
exclude functionals with $m^*=m$ as we have demonstrated in this 
work. The fact that our results comply with the criteria of \cite
{stone} is an indication that nuclear density functionals should be 
based as much as possible on a solid microscopic input. In this 
respect it is to be noted that out of the five functionals which 
survived the criteria in \cite{stone}, three are also based on a 
microscopic input.

\begin{acknowledgments}
We are grateful to P. Danielewicz for useful discussions. We would 
like to thank N. Schunck for a careful reading of the manuscript and 
many useful suggestions. Work supported in part by MICINN grants 
Nos. FPA2009-08958, FIS2009-07277, and FPA2008-03865-E/IN2P3 and by 
the Consolider-Ingenio 2010 program CPAN CSD2007-00042 and MULTIDARK 
CSD2009-00064. X. V. also acknowledges the support from 
FIS2008-01661 (Spain and FEDER) and 2009SGR-1289 (Spain). Support by 
CompStar, a Research Networking Programme of the European Science 
Foundation is also acknowledged.
\end{acknowledgments}

\appendix
\section{Sum rule approach to giant monopole and quadrupole resonances}

In this Appendix we briefly summarize the basic theoretical aspects of 
the sum rule approach to obtain moments of the RPA strength functions.
For more details we refer the reader to Refs. \cite{RS80,boh79,pat02,cen05,sou04}.

In the sum rule approach it is enough to know few low energy weighted
moments of $S(E)$ (the so-called sum rules) to estimate the excitation 
energies of the giant resonances. The sum rules are defined as
\beq
m_k = \sum_n E_n^k {| \langle n |Q| 0 \rangle |}^2 ,
\label{eq01} 
\eeq
where $Q$ is the operator associated to the excitation, $|0>$ and
$|n>$ are the exact ground
and excited states respectively and $E_n$ the excitation energies.
If $k$ is an odd integer, the sum rules $m_k$ can be computed as
the expectation values in the ground state $|0\rangle$ of some
commutators which involve the hamiltonian $H$ and the operator $Q$ 
(assumed to be a hermitian and one-body operator) \cite{boh79}, as for example
\beq
m_1 = \langle 0 | \big[ Q,\big[ H,Q \big] \big] | 0 \rangle
\label{eq06} 
\eeq
and
\beq
m_3 = \langle 0 | \big[ \big[ Q,H \big], \big[ H, \big[ H,Q \big] 
\big] \big]| 0 \rangle,
\label{eq07} 
\eeq
from where an average excitation energy can be estimated as
\beq
\bar{E_3} = \sqrt{\frac{m_{3}}{m_{1}}}.
\label{eq04} 
\eeq

The full calculation of the sum rules is still a complicated task
because the exact ground-state wavefunction is, in general, unknown.
However, if the moments are computed within the 1p1h RPA, it is
possible to replace the exact ground-state wavefunction by the
uncorrelated HF one to obtain the sum rules (\ref{eq06})
and (\ref{eq07}) \cite{boh79}.
In spite of all these simplifications, the calculation of the sum
rules is only easy to handle in few particular cases, namely
the ISGMR and ISGQR ones. In these cases the sum rules needed to estimate
the average excitation energy of these resonances can be simplified
as we will discuss below.

For the isoscalar monopole and quadrupole oscillations the
corresponding excitation operators are taken as              
\beq
Q_M = \sum_{i=1}^A r_i^2 \qquad Q_Q = \sum_{i=1}^A (r_i^2 -3 z_i^2).
\label{eq02}
\eeq
The underlying effective interaction associated to the BCPM energy
density functional commutes with the operator $Q$
and, therefore, the only contribution to the commutator $[Q,[H,Q]]$ 
in eq. (\ref{eq06}) comes
from the kinetic energy operator. Consequently \cite{RS80,boh79}
\beq
m_1 = \frac{2 \hbar^2}{m} A \langle r^2 \rangle \quad (L=0) \qquad
m_1 = \frac{4 \hbar^2}{m} A \langle r^2 \rangle \quad (L=2),
\label{eq09b} \eeq
where the expectation value of the operator $r^2$ is calculated
with the HF wave functions.

The direct evaluation of the commutators entering in Eq. (\ref{eq07})
to compute the $m_3$ sum rule is a rather cumbersome task, that in the
case of the monopole and quadrupole oscillations can be avoided with
the help the so-called scaling method \cite{boh79}. We start from the
scaled ground-state wave function that in the case of the monopole reads
\beq                                                      
\phi_{\lambda}^M = \lambda^{3 /2}
\phi_0(\lambda x, \lambda y, \lambda z),
\label{eq2} \eeq
where $\lambda$ is an arbitrary scaling parameter.
In this case the $m_3$ sum rule can be expressed by means of the
second derivative of the scaled ground-state energy \cite{boh79}, 
that is,
\beq
m_3 = \frac{1}{2} \frac{\partial^2}{\partial \lambda^2}
\big[\langle \Phi_{\lambda} \vert H \vert \Phi_{\lambda} \rangle 
\big]_{\lambda=0} .
\label{eq3} \eeq

In the scaling approach the BCPM energy density functional reads
\beq
E(\lambda) = T(\lambda) + E^{\infty}_{int}(\lambda)
+ E^{FR}_{int}(\lambda) + E_C(\lambda) + E^{s.o}(\lambda),
\label{eq6a} \eeq
where the scaled kinetic, spin-orbit and Coulomb energies can be found 
in Ref. \cite{boh79}.
The scaled bulk contribution can be easily obtained from the scaling
of the particle density ($\rho_{\lambda}(\vect{r})= \lambda^3 \rho(\lambda \vect{r})$). 
The scaling of the finite range contribution
has been described in Refs. \cite{pat02,sou04}. 
It is easy to show that the scaling of the Hartee term reduces to a
renormalization of the range $\mu$ of the interaction, or in
other words that the scaled relative coordinate is
$s_{M,\lambda}= \vert \vec{r} - \vec{r'} \vert/\lambda$.
In this way, the second derivative with respect to the scaling parameter
$\lambda$, needed to compute the $m_3$ sum rule, finally reads \cite{pat02}:
\beq
\frac{d^2 E^H_M(\lambda)}{d \lambda^2} \vert_{\lambda=1} =
\frac{1}{2} \iint \! d \vec{r} d \vec{r'} \rho(\vec{r})
\rho(\vec{r'}) \left[ 2 s \frac{dv}{ds} + s^2 \frac{d^2 v}{d s^2} \right]
\label{eq6d} 
\eeq
In the BCPM energy density functional, the form factor $v(s)$ is of Gaussian 
type with a range $\mu$. In this case
the second derivative of the scaled finite range contribution can be 
recast as \cite{sou04}
\beq
\frac{d^2 E^H_M(\lambda)}{d \lambda^2} \vert_{\lambda=1} =
\mu^2 \frac{d^2 E^H_M(\mu)}{d \mu^2}.
\label{eq6e} 
\eeq
Finally the  $m_3$ for the monopole oscillation computed with the BCPM 
energy density functional can be written as
 \begin{eqnarray}
 m_3(L=0) & = &
 \frac{1}{2} \left(\frac{2 \hbar^2}{m}\right)^2  \left[ 2T + 20 E^{s.o}
 + \frac{d^2 E^{\infty}(\lambda \rho)}{d \lambda^2}
\vert_{\lambda=1} \right. \nonumber \\
& + &  \left.
\mu^2 \frac{d^2 E^{FR}_M(\mu)}{d \mu^2}\right]
\end{eqnarray}

In the case of the quadrupole oscillation the scaling
transformation of each single-particle wavefunction is given by
\cite{boh79}

\beq
\rho_{\lambda}(\vec{r}) = \rho(\lambda x, \lambda y, z^2/\lambda^2).
\label{eq4c} 
\eeq

Notice that under this transformation the volume element is 
conserved because of the surface nature of the quadrupole oscillation.

As explained in detail in Ref. \cite{pat02}, a pure Hartree term scales in the 
quadrupole case as
\beq
E^H_Q(\lambda) = \frac{1}{2} \iint \! d \vec{r} d \vec{r'}
\rho(\vec{r}) \rho(\vec{r'}) v(s_{Q,\lambda}),
\label{eq7d} 
\eeq
where $s_{Q,\lambda}= (s_x/\lambda, s_y/\lambda, \lambda^2 s_z)$. After some 
algebra the surface contribution to the $m_3$ sum rule in the scaling approach 
for the quadrupole oscillation can be written as (see Ref.\cite{pat02} for 
details) 
\beq
\frac{d^2 E^H_Q(\lambda)}{d \lambda^2} \vert_{\lambda=1} =
\frac{2}{5} \iint \! d \vec{r} d \vec{r'} \rho(\vec{r})
\rho(\vec{r'}) \left[ 4 s \frac{dv}{ds} + s^2 \frac{d^2 v}{d s^2} \right],
\eeq
that in the case of the BCPM functional owing to the gaussian character of 
the form factor $v$  can also be written as:
\beq
\frac{d^2 E^H_Q(\lambda)}{d \lambda^2} \vert_{\lambda=1} =
\frac{4}{5} \left[ \mu^2 \frac{d^2 E^H(\mu)}{d \mu^2} -
2 \mu \frac{d E^H_Q (\mu)}{d \mu} \right].
\eeq

This is due to the fact of the volume conservation in the quadrupole 
scaling, the bulk interaction part of the BCP energy density 
functional does not contribute to the $m_3$ sum rule in this case. 
The contribution from kinetic and spin-orbit energies can be easily 
obtained from the scaling transformation of the single-particle wave 
functions assuming a spherically symmetric ground-state \cite
{boh79}. The Coulomb contribution can also be obtained from Eq. (\ref
{eq7d}) using the Coulomb form factor $e^2/s$ \cite{pat02}. 
Collecting all these contributions, the $m_3$ sum rule for the 
quadrupole oscillation can finally be written as
 \begin{eqnarray}
 m_3(L=2) & = &
 4\left(\frac{2 \hbar^2}{m}\right)^2  \left[ T + \frac{1}{4} E^{s.o}
 - \frac{1}{5} E_C \right. \nonumber \\
& + & \left. \frac{1}{10} \left[ \mu^2 \frac{d^2 E^H_Q(\mu)}{d \mu^2}
- 2 \mu \frac{d E^H_Q(\mu)}{d \mu} \right] \right].
\end{eqnarray}

Once the $m_3$ sum rule has been obtained through the scaling 
method, an estimate of the excitation energy of the ISGMR and ISGQR 
can be calculated with the help of Eqs. (\ref{eq04}) and(\ref{eq09b})
\beq
E^S_M = \sqrt{\frac{m_3}{m_1}} .
\label{eq19a} \eeq

Let us now consider a nucleus, described by a Hamiltonian $H$, under the 
action of a weak one-body field $\eta Q$. Assuming $\eta$ sufficiently 
small, so that the perturbation theory holds, the variation of the 
expectation value of $Q$ and $H$ are directly related to the $m_{-1}$ 
moment \cite{boh79},
\beq
m_{-1} = \sum_n \frac{{|\langle n|Q|0\rangle|}^2}{E_n} =
- \frac{1}{2} \bigg[ \frac{\partial \langle Q \rangle}{\partial \eta} \bigg]_{\eta=0} =
 \frac{1}{2} \bigg[ \frac{\partial^2 \langle H \rangle}{\partial \eta^2} 
\bigg]_{\eta=0}.
\label{eq12b} \eeq 

Therefore at RPA level the average energy of the giant resonances can also be 
estimated performing constrained HF calculations, i.e. looking for the 
HF solutions of the constrained Hamiltonian
\beq
H(\eta) = H - \eta Q
\label{eq17} 
\eeq
where $Q$ is the collective monopole (quadrupole) operator defined previously 
(see Eq. (\ref{eq02}). From the constrained HF wave-function $\Phi(\eta)$  
solution of (\ref{eq17}), the RPA $m_{-1}$ moment can be written as
\begin{eqnarray}\label{eqsumr2}
 m_{-1} & = & - \frac{1}{2}
 \left[ \frac{\partial}{\partial \eta}
 \langle \Phi(\eta)\vert Q \vert \Phi(\eta) \rangle \right]_{\eta=0} \\ \nonumber
& = & \frac{1}{2}
\left[ \frac{\partial^2}{\partial \eta^2}
\langle \Phi(\eta)\vert H \vert \Phi(\eta) \rangle \right]_{\eta=0},
\end{eqnarray}
from where another estimate of the average energy of the isoscalar 
giant resonances is given by
\beq
\bar{E}_1 = \sqrt{\frac{m_1}{m_{-1}}}.
\label{eq19} \eeq

Due to the fact that the RPA moments fulfill $\sqrt{m_3/m_1} \ge 
m_1/m_0 \ge \sqrt{m_1/m_{-1}}$ \cite{boh79}, the average energies 
$\bar{E}_3$ and $\bar{E}_1$ values are an upper and lower bound of 
the mean energy of the resonance and their difference is related 
with the variance of the strength function (resonance width)
\beq
\sigma = \frac{1}{2} \sqrt{ \bar{E_3}^2 - \bar{E_1}^2}.
\label{eq21} \eeq

\end{document}